\pdfoutput=1
\documentclass{article}

\usepackage{PRIMEarxiv}

\usepackage[utf8]{inputenc} 
\usepackage[T1]{fontenc}    
\usepackage{hyperref}       
\usepackage{url}            
\usepackage{booktabs}       
\usepackage{amsfonts}       
\usepackage{nicefrac}       
\usepackage{microtype}      
\usepackage{lipsum}
\usepackage{float}
\usepackage{pifont}
\usepackage{verbatim}
\usepackage{fancyhdr}       
\usepackage{graphicx, subcaption}
\usepackage{multicol}
\usepackage{amsmath, amsthm}
\usepackage{makecell}
\usepackage[ruled]{algorithm2e}
\usepackage[labelfont=bf]{caption}
\usepackage[marginal]{footmisc}

\graphicspath{{media/}}     

\theoremstyle{definition}
\newtheorem{rul}{Rule}[section]
\newtheorem{lemma}{Lemma}[section]
\newtheorem{proo}{Proof}[section]
\newtheorem{example}{Example}[section]

\title{B-rep Boolean Resulting Model Repair by Correcting Intersection Edges Based on Inference Procedure
}

\author{
  Haomian Huang*, Li Chen, Enya Shen, Jianmin Wang\\
  School of Software\\
  Tsinghua University \\
  Beijing, China\\
}

\begin{document}
\footnotetext{\texttt{email:huanghm21@mails.tsinghua.edu.cn}}
\maketitle
\begin{abstract}
As the most essential part of CAD modeling operations, boolean operations on B-rep CAD models often suffer from errors. Errors caused by geometric precision or numerical uncertainty are hard to eliminate. They will reduce the reliability of boolean operations and damage the integrity of the resulting models. And it is difficult to repair false boolean resulting models damaged by errors. In practice, we find that the illegal boolean resulting models stem from the false intersection edges caused by errors. Therefore, this paper proposes an automatic method based on set reasoning to repair flawed structures of the boolean resulting models by correcting their topological intersection edges. We provide a local adaptive tolerance estimation method for each intersection edge based on its geometric features as well as its origin. Then, we propose a set of inference mechanisms based on set operations to infer whether a repair is needed based on the tolerance value and how to correct the inaccurate intersection edge. Our inference strategies are strictly proven, ensuring the reliability and robustness of the repair process. The inference process will transform the problem into a geometric equivalent form less susceptible to errors to get a more accurate intersection edge. Since our inference procedure focuses on topological features, our method can repair the flawed boolean resulting models, no matter what source of errors causes the problem. 
\end{abstract}


\begin{multicols}{2}
\section{Introduction}
\label{1}
Model problems caused by errors are common in the process of boolean operations in CAD kernels. Based on their sources, there are four main kinds of errors\cite{fang1991robustness}: inaccurate input data, approximation errors, floating-point arithmetic errors, and inherent accuracy limitations of the geometric intersection algorithms. If the CAD kernels cannot correctly deal with these errors, then the topology of the resulting models may be corrupted. It will generate some flawed structures that users do not expect. Furthermore, models with topology errors may cause the corruption of the program when users do some subsequent operations on them. Therefore, without a solution to error problems, CAD software cannot be applied in practice.

To tackle these error problems, many researchers have done a lot of work. One solution is to implement a more accurate geometric intersection algorithm. 
There are many numerical methods to solve polynomial systems that arise in geometric intersections. Newton's method\cite{nocedal1999numerical} is simple to implement, but it cannot find all solutions and always gets stuck in local optimums. There are some methods for solving the global optimization problem in polynomial systems\cite{lasserre2001global}, but these methods lack accuracy and robustness. To overcome these disadvantages, researchers proposed subdivision methods to properly divide the intersection areas into multiple regions and then use different strategies to find the solutions based on each region's conditions\cite{shao2022robust}. However, all these improved intersection algorithms only reduce the latter two kinds of errors. Besides, they do not enhance CAD kernels' ability to deal with the accumulating errors, which will break models after a sequence of operations. Another way to deal with error problems is tolerant modeling, which enables CAD software to accept less precise data as input. It allows every vertex and edge to have its tolerance value. These local tolerance values can heal the flawed structures caused by errors like leaky and disconnection while maintaining the accuracy and integrity of the geometric shapes. But setting appropriate tolerance values can be a difficult task. And improper local tolerance values will cause inconsistencies at touch areas between two different models when assembling or self-intersection inside the topological entity, corrupting the whole model. What is worse, the tolerance values will bring uncertainty and inconsistency to the boolean operations, which generate flawed structures in the resulting models.

With the development of artificial intelligence, neural networks have been used to process CAD models. However, due to the complexity and ambiguity of B-rep models, neural networks nowadays can only handle some elementary tasks like segmentation and classification, and even then, there is still a high chance of misjudgment, which is unacceptable in practice. CAD kernels require that the algorithms must be highly accurate and stable. Otherwise, a small error will lead to unexpected wrong results.

Consequently, the solution to error problems must ensure the correctness of the resulting models regardless of the source of errors. Moreover, the method must be highly accurate and stable. Only then can this method be embedded into the CAD kernel and actually used in practice. According to our hundreds of thousands of tests on the common shapes of industrial parts, we find that the main reason why boolean operations fail is the intersection inconsistency caused by errors. Therefore, in this paper, we focus on repairing boolean resulting models with flawed structures by correcting the inaccurate intersection edges. Based on the topology level of the intersection inconsistency, we classify it into three types: face-face, edge-face, and edge-edge. For each type of scenario, a well-designed inference procedure is proposed to detect and repair this kind of inconsistency. All inference procedures are strictly proven and will help us avoid misjudgment, which will significantly enhance the accuracy and stability of the repair method. Our method first calculates appropriate tolerance values for every intersection edge. Then, we combine the intersection edges into the structure of the intersection graph, introduced in \textbf{section \ref{subsection:Intersection Graph}}. Then, we use the tolerance values of each edge and the intersection graph to direct the inference procedures, which will only correct inaccurate intersection edges that will cause flawed structures in the resulting models. This method has the following contributions:

\noindent\textbullet Proposing a framework for repairing boolean resulting models by correcting intersection edges. In previous work, some researchers have noticed the intersection inconsistency in
polyhedral models\cite{hoffmann1989robust}, but to the best of our knowledge, we are the first to improve the boolean results of parametric B-rep models based on their intersection edges. 

\noindent\textbullet Organizing intersection edges into an intersection graph, which carries much geometric and topological information about intersection edges and the resulting model, to help make a reasonable repair. Compared with extracting information from resulting models,  information from intersection edges is more intuitive, and the process is more efficient due to the simple topological structure of intersection graphs. With the help of intersection graphs, we can preserve the tiny features of the resulting models very well in the repair process.


\noindent\textbullet Valid inference procedures which use topological features to detect and correct inaccurate intersection edges. Our inference process will locate the problem intersection edges and fix them with more accurate results. This process is directed by the feature-based tolerance values of intersection edges, which are calculated automatically. Therefore, it is very friendly for users. Besides, Our inference strategies are proved by set reasoning, which promises the soundness of our algorithms.

\section{Related Work}

\noindent\textbf{Robust boolean operations on plane-based models} 
A considerable part of boolean operation errors come from the process of intersection calculation. In plane-based models like mesh, there are several methods to reduce the errors. An implicit plane-based boundary representation\cite{bernstein2009fast} is proposed to solve this problem. In this method, each point is implicitly represented as the intersection of three planes, and each straight line is implicitly represented as the intersection of two planes. Besides, a technology called filtered floating point predicate\cite{richard1997adaptive} is introduced to reduce float errors. However, this method will hurt the efficiency of the whole process. Therefore, homogeneous coordinates\cite{nehring2021fast} are used to totally avoid float errors. By this method, all float point operations are converted into integer operations, but using homogeneous coordinates will make some algorithms difficult to implement, such as the triangulation of mesh grids. That is why this method cannot be widely used. And in previous work, researchers had noticed that the intersection inconsistency between two models was the main reason why boolean operations failed. Therefore, they proposed robust boolean operations on polyhedral solids based on cross-section graphs and implicit representation\cite{hoffmann1989robust}. A cross-section graph comprises the intersection of a solid with a face. Edges in the graph are well-oriented to partition the areas of the face as being inside, on, or outside the solid. Besides, they perform robust incidence tests to help classify the areas. By using the implicit representation of points and lines, an incidence test can be transformed into multiple point-on-plane problems, which will enhance the accuracy. However, these above methods can only be applied on straight lines and planes. In industrial design, B-rep models based on parametric curves and surfaces are more widely used than polyhedral models. Unlike mesh or polyhedral models, whose errors mainly come from float errors, parametric B-rep models' error composition is more complex. Complex curves and surfaces need to be fitted with splines, which bring approximation errors after the formation of B-rep models. Moreover, the intersection of curves and surfaces is more complicated than that of straight lines and planes in mesh models. The intersection algorithms have calculation errors except for float errors, which will continue accumulating during the modeling process. Therefore, tolerant modeling technology\cite{jackson1995boundary} is proposed to solve the problem of model illegality caused by errors.

\noindent\textbf{Tolerant modeling} 
Generally, precise models are best for users. But for the above reasons, it is hard to get accurate data in some situations. Tolerant modeling allows the program to accept less accurate data while keeping the validity of topology structures\cite{acis2023tolerant}. Mainstream CAD kernels have multiple kinds of global tolerance variables. For example, the positional tolerance value is the smallest distance between two different topological vertices, and the fitting tolerance value is used to guide the accuracy of fitting algorithms for complex curves and surfaces. The global tolerance method is efficient, and it can handle most situations. But more is needed to solve some exceptional cases. And it will be difficult to transfer models' data from one CAD kernel to another with different tolerance settings. To avoid these disadvantages, local tolerant modeling is proposed\cite{jackson1995boundary}. It allows basic topology structures such as edges and vertices to have their own tolerance values, which does not impact the robustness and reliability of the precise modeling while strongly enhancing the ability to accept less accurate data to the model. 

\noindent\textbf{Tolerant analysis} 
Determining suitable tolerance values for topology structures and whether present tolerance values can satisfy following operations become a crucial topic in tolerant modeling\cite{chen2014comprehensive}.  Tolerance-map(T-map)\cite{davidson2002new} is a hypothetical Euclidean-Space point set representing all probabilities of a feature in its tolerance zone. When assembling two components with the same features, the tolerance propagates by calculating the Minkowski sum of two T-maps. After this method was proposed, T-maps were developed for other features quickly\cite{mujezinovic2004new}. However, the T-maps of some kinds of features are high dimensional, which makes the calculation of the Minkowski sum inefficient\cite{delos2021polyhedral}. The matrix model\cite{desrochers1997matrix} uses a displacement matrix to describe the small displacements of a feature in its tolerance zone\cite{herve1978analyse}. Therefore, all points of this feature can be bounded by the inequality resulting from the combination of displacement matrix and tolerance values. The calculation can be very efficient, but different choices of the point may lead to different results\cite{whitney1994representation}, which will cause some ambiguities when tolerance propagates between different models. 
 
\noindent\textbf{Repair directly on resulting models}
Repairing flawed structures caused by errors in resulting models is another way to handle error problems. In the exact B-rep repairing method, models are checked and corrected by analyzing their topological and geometric elements\cite{hoffman1998cad}. And in the faceted B-rep repairing method, original models are approximated as polyhedrons, making the check and repair process more efficient\cite{barequet1997using}. However, errors during the previous process may make the resulting models totally collapse, which means these methods can only correct limited kinds of errors. Therefore, a repair method based on design history is proposed\cite{yang2006repairing}. In some CAD kernels, model data includes not only its topological and geometric elements but also its design history, which tells the users how this model is constructed. Consequently, even if the resulting model collapses due to some errors, the program can still reconstruct the model partly and detect where the errors occur. However, this method depends on the definitions of error types, which may falsely modify the flawless models. Besides, analyzing design history data is time-consuming, and some CAD systems do not have such information for each model.

\noindent\textbf{Deep learning} 
Recently, neural networks have been used to process CAD B-rep models. Compared with mesh and point cloud data, B-rep models formed by parametric curves and surfaces are much more complex, and few neural networks can consume such B-rep data directly. To overcome the difficulty, UV-Net\cite{jayaraman2021uv} extracts geometric features from the parameter domain as regular grids(UV-grids) and uses a face-adjacency graph representing the topological connection to form the structure of the graph neural network. Then, image convolutions on the curve and surface UV-grids are performed to get their hidden features. The hidden features, which are the output of the CNNs, are then treated as input edge and node features to the graph neural network. BRepNet\cite{lambourne2021brepnet}, however, uses convolution networks only. The input of the convolution networks is the feature vectors that encode the geometric information of the faces, edges, and coedges. Compared with UV-grids, feature vectors are much simpler and only include some basic information like the geometric types, the area of a face, and the length of an edge. Therefore, these features can be extracted very efficiently. And the convolutional kernels are well-designed based on the index array of topological entities and their topological connections, which enables the hidden features to transfer from one entity to its neighbor entity. The output of the above networks can achieve satisfactory results in several downstream applications like segmentation and classification. However, due to the complexity and ambiguity of  B-rep models, no neural network currently can detect the flawed structures of boolean resulting models, let alone repair them. 

In our method, we detect errors before forming resulting models in boolean operations by analyzing the intersection edges, and we only use the topological and geometric information of the models that every CAD kernel will provide for users. Besides, our method does not need to improve the geometric intersection algorithm, which will make the algorithm easy to apply in other CAD kernels. Moreover, the strict inference process ensures our method does not modify any flawless structures of models.

\section{Background Knowledge}
\subsection{Boolean operations}
As the most essential part of the CAD kernel, boolean operations greatly suffer from error problems. Boolean operations are used to construct complex models based on basic models like cubes, cones, and torus. Each boolean operation of two models can be divided into four stages:

\noindent\ding{172} Intersection. Calculate the face-face intersection of each face of the two models separately, and get all the intersection edges. 

\noindent\ding{173} Imprint. Imprint all these intersection edges onto both models and split each face of the two bodies. After this step, all faces of the two models fit the following property--the face of a model either lies totally inside the body of the other model, or totally outside, or totally on the boundary of the other model. 

\noindent\ding{174} Classification. Determine whether each face should be kept or discarded based on the input boolean operator. 

\noindent\ding{175} Merge. Merge all the remaining faces into one body as the final result of the boolean operations, and correctly maintain its topology tree to ensure the resulting model's topology is legal.  
\subsection{Geometric Intersection and Topological Intersection}

In the boundary representation(B-rep) model, the shape of an entity is decided by its geometric element, and its boundary is composed of lower-dimension entities. For example, faces are bounded by edges, and edges are bounded by vertices. Faces, edges, and vertices are direct compositions of a model, and we call them topological structures because they will connect with other topological structures to form a model. And the geometric element of a face, an edge, or a vertex, we call it geometric structure. A geometric structure will not connect with any other geometric structure. It is a mathematical description of a shape. To distinguish them, we call the geometric structure of a face a surface, the geometric structure of an edge a curve, and the geometric structure of a vertex a point(position). Therefore, there are two kinds of intersection ways: geometric and topological.

Geometric intersections will not consider any boundary information of its corresponding topological entity. It is the computation that happens between geometry structures.

Topological intersections, as the name indicates, will consider the boundary of the topological entity. It will filter the results of geometric intersections and leave the valid parts inside the boundary of the corresponding topological entities. The process of topological intersection runs following the order from high dimensions to low dimensions. Generally, a face-face intersection process will run multiple times edge-face intersection processes to get boundary information of each face. So will the edge-face intersection process run multiple times edge-edge intersection. Because high-dimension topological intersections will use the information returned from lower-dimension topological intersections, in this process, errors will accumulate. Therefore, the accuracy of higher dimension topological intersection will be lower.

\subsection{Topology Set Computation and Inference Method}

We introduce the inference method based on the already known topological information to determine whether intersection results need to be fixed. In order to describe the inference process better, we specify the following topological set notation:

\begin{center}
    \begin{tabular}{cl}
    \centering
        $a$ &  \makecell[l]{Set containing the topological entity $a$\\ itself and all its sub topological structu\\-res}\\
        $a \cap b$&\makecell[l]{Set containing the topological intersect\\-ion result of entity $a$ and entity $b$}\\
        $a\in b$& \makecell[l]{Entity $a$ is a sub topological structure of\\ entity $b$}\\
    \end{tabular}
\end{center}

For example, face $f$ represents not only the face itself but contains all sub-topological structures, including all its boundary edges and all vertices of these edges. $e\cap f$ is the topological entity set containing the topological intersection result of edge $e$ and face $f$. If $e$ is a boundary edge of $f$, then $e\in f$. 

Obviously, topology set computation satisfies commutative and associative laws, that is:

$$a\cap b = b\cap a$$
$$a\cap b\cap c = a\cap(b\cap c)$$

Besides, there are two extra inference rules based on the feature of topological entities and topological intersections:

\begin{rul}
\label{rule1}
    $e\in f\Rightarrow e\cap f = e$
\end{rul}

That is, if edge $e$ is a boundary edge of face $f$, then the intersection result of $e$ and $f$ is $e$ itself.

\begin{rul}
\label{rule2}
    $v\in e\cap f\Rightarrow v=e\cap f$, in a sufficiently small neighborhood of $v$.
\end{rul}

That is, if vertex $v$ is one of the intersection resulting entities of $e$ and $f$, then $v$ is the only intersection result of $e$ and $f$ within a sufficiently small neighborhood of $v$.

Using these two rules, we can have the following lemma:
\begin{lemma}
\label{lemma1}
$e_1\in f_1, e_2\in f_2, e_1\cap e_2 = v_0\Rightarrow f_1\cap e_2=v_0, e_1\cap f_2=v_0$, in a sufficiently small neighborhood of $v_0$
\end{lemma}

The proof process is shown as follows:
\begin{proo}
Using \textbf{Rule \ref{rule1}}, we know that 
\begin{align*}
    &e_1\in f_1\Rightarrow e_1\cap f_1 = e_1 \\
\end{align*}
Therefore,
\begin{align*}
    &v_0 = e_1\cap e_2 = e_1\cap f_1\cap f_2\Rightarrow v_0\in e_1\cap f_2&\\
\end{align*}
Using \textbf{Rule \ref{rule2}}, in a sufficiently small neighborhood of $v_0$, we know that
\begin{align*}
    &v_0\in e_1\cap f_2\Rightarrow v_0=e_1\cap f_2
\end{align*}
Similarly, we can get $v_0=f_1\cap e_2$ in a sufficiently small neighborhood of $v_0$.
\qed
\end{proo}

This lemma will help us to simplify problems in some situations.

\section{Method}
\label{2}
\subsection{Feature-based Tolerance Value for Intersection Edges}
To use \textbf{Lemma \ref{lemma1}}, we should run our inference procedure in a local area. Therefore, we should calculate a proper tolerance value for each intersection edge to direct the inference process. Global tolerance can handle many situations. However, in our experiments, we find that using a global tolerance value will ignore fixing some flawed cases, especially when the shape of an intersection edge is complex. Consequently, we need to examine the features of each intersection edge and calculate an appropriate local tolerance value for it. 

Intersection edges with different geometry have different accuracy. For instance, the error of a straight line is always much smaller than that of a high-degree spline curve. Therefore, the higher the degree of the mathematical expression of the geometry curve of an intersection edge, the larger its tolerance value should be.

Besides an intersection edge's geometry, the shapes of two faces that generate the intersection edge are also important factors. A straight line generated by two planes is more precise than two spline surfaces. Consequently, intersection edges coming from two faces with higher-order surfaces should have larger tolerance values.

Because different geometric intersection algorithms have different accuracy when dealing with different situations, it will take much work to determine an accurate tolerance value for an intersection edge in different situations. In practice, we find the tolerance values that can be used to direct inference procedure are positively correlated with several factors, so we have the following simple empirical formula to calculate a feature-based tolerance value $t$ for an intersection edge $e_0$ generated by the intersection of face $f_1$ and $f_2$:
$$t=k\times \textrm{d}(f_1)\times\textrm{d}(f_2)\times\textrm{d}(e_0)\times t_0$$

In the above formula, $\textrm{d}$ means the maximum degree of the mathematical expression of the edge or face's geometry. $t_0$ means the default global positional tolerance value. Since errors may accumulate during the calculation process, we connect these coefficients using multiplication signs. Therefore, in the most straightforward situation, two intersecting planes making a straight line, the default tolerance value is significant enough to direct the inference process when $k$ equals 1. And the coefficient $k$ describes the level of the model's geometry deviating from the actual position.  In some situations, the geometry of the faces or edges may be represented by fitting spline. And the fitting spline has its inherent numerical inaccuracy. Therefore,  $k$ is defined by the following formula:

$$k = \textrm{maximum}{(1, \frac{t_{f_1}}{t_0},\frac{t_{f_2}}{t_0},\frac{t_{e_0}}{t_0})}$$

In the above formula, $t_{f_1}$, $t_{f_2}$ and $t_{e_0}$ are separately the fitting errors of $f_1$, $f_2$ and $e_0$. Because fitting errors are always much larger than the global positional tolerance value, their ratios can reach a value of more than 1000. Consequently, considering only the largest ratio is enough for us to achieve a large enough $k$ value.


\subsection{Intersection Graph}
\label{subsection:Intersection Graph}
The intersection graph is a topological structure consisting of intersection edges and their vertices, which describes the intersection result of two models. The intersection edges with vertices in the same position will be connected, and these vertices in the same position will be merged into one. 

\begin{figure}[H]
    \centering
    \subcaptionbox{}{
        \includegraphics[width = .3\linewidth]{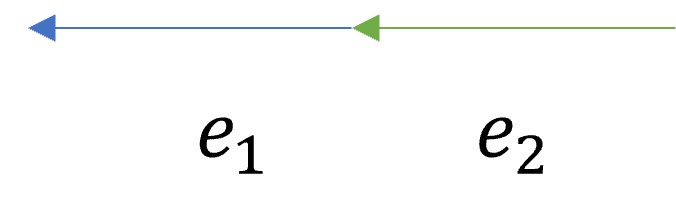}
    }
    \subcaptionbox{}{
        \includegraphics[width = .3\linewidth]{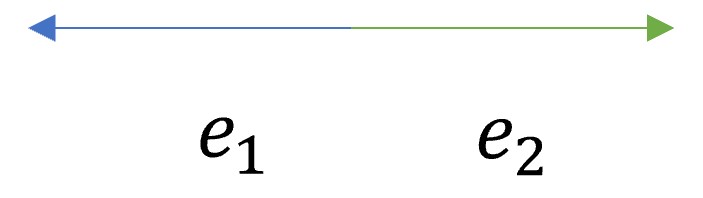}
    }
    \subcaptionbox{}{
        \includegraphics[width = .3\linewidth]{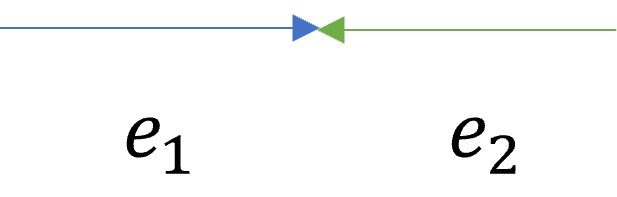}
    }
    \caption{(a) two edges are connected head to tail, $e_1$ is $e_2$'s successor and $e_2$ is $e_1$'s predecessor. (b) two edges are connected tail to tail, $e_1$ is $e_2$'s predecessor and $e_2$ is $e_1$'s predecessor. (c)two edges are connected head to head, $e_1$ is $e_2$'s successor and $e_2$ is $e_1$'s successor.}
    \label{fig:three connected ways}
\end{figure}

\begin{figure}[H]
    \centering
    \includegraphics[scale=0.9]{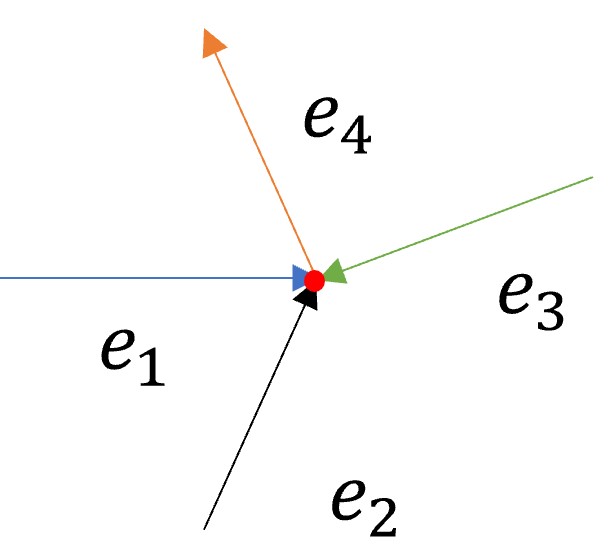}
    \caption{Four edges share one vertex and make a branch at this vertex. Based on their directions, there is a legal connection ways that $e_1$'s successor is $e_2$, $e_2$'s successor is $e_3$, $e_3$'s successor is $e_4$, and $e_4$'s predecessor is $e_1$. The legal ways to connect these edges are not unique. There is another way that $e_1$'s successor is $e_3$, $e_3$'s successor is $e_4$, $e_4$'s predecessor is $e_2$, and $e_2$'s successor is $e_1$.}
    \label{fig:branch}
\end{figure}

In the intersection graph, each edge has a predecessor and a successor based on its own direction, pointing from its start vertex to its end vertex. However, two neighbor edges in the graph do not need to be connected head to tail, which means their relative orientation can be arbitrary. Shown in Fig.\ref{fig:three connected ways},  the predecessor and successor of an edge must follow the rule that the predecessor is one of the neighbor edges that is connected to its start vertex, and the successor is one of the neighbor edges that connected to its end vertex. This design allows the intersection graph to have branches, shown in Fig.\ref{fig:branch}. When branching at a vertex, all edges sharing this vertex will be organized like a circular linked list, and the order of the edges in this list is not unique.

Specially, the predecessor of an edge will be itself when its start vertex does not connect to other edges. And similarly, if the end vertex of an edge does not connect to other edges, the successor will be itself. This feature will make adding or deleting an edge in the intersection graph more convenient since we do not need to distinguish whether this edge connects to other edges. 

Although intersection graphs' topological structures are quite simple, they carry much geometric and topological information about the resulting models.  Compared with extracting features from the resulting models, the information from intersection graphs is more intuitive and accurate since the extracting steps may bring ambiguity or ignore some small but important features. With the information given by intersection graphs, our inference procedures can be more robust and valid.

\subsection{Inference Algorithms}

In the repair process, we must ensure that we will not modify any accurate edges while calculating a more precise result for the flawed intersection edges. And our inference algorithms can satisfy this by using the following three strategies in different situations:

\noindent\ding{172} Transform a high-dimension intersection problem into a low-dimension intersection problem to get a more accurate intersection result since high-dimension topological intersections are more affected by accumulating errors than low-dimension topological intersections.

\noindent\ding{173} Use the intersection graph to get more information from other parts of the model, which can help avoid misjudging. Intersection graphs can help us identify some small features that are easily mistaken for flawed structures generated by errors.

\noindent\ding{174} Use the information we already know to substitute the intermediate result, which can help get a more accurate final result. Because of the accumulation of errors, using the intermediate result in computation will further reduce accuracy.

As mentioned earlier in this paper, the main reason why boolean operations fail is intersection inconsistencies caused by errors. And there are three types of intersection inconsistency classified by their topology level: face-face, edge-face, and edge-edge. In the following part, we will separately discuss three topological manifestations of the three types of inconsistency and how our inference algorithms detect and repair them.

\subsubsection{Disconnection of two Intersection Edges that Should Be Connected}
\label{subsubsection:Disconnection}
\begin{figure}[H]
    \centering
    \includegraphics[scale=0.8]{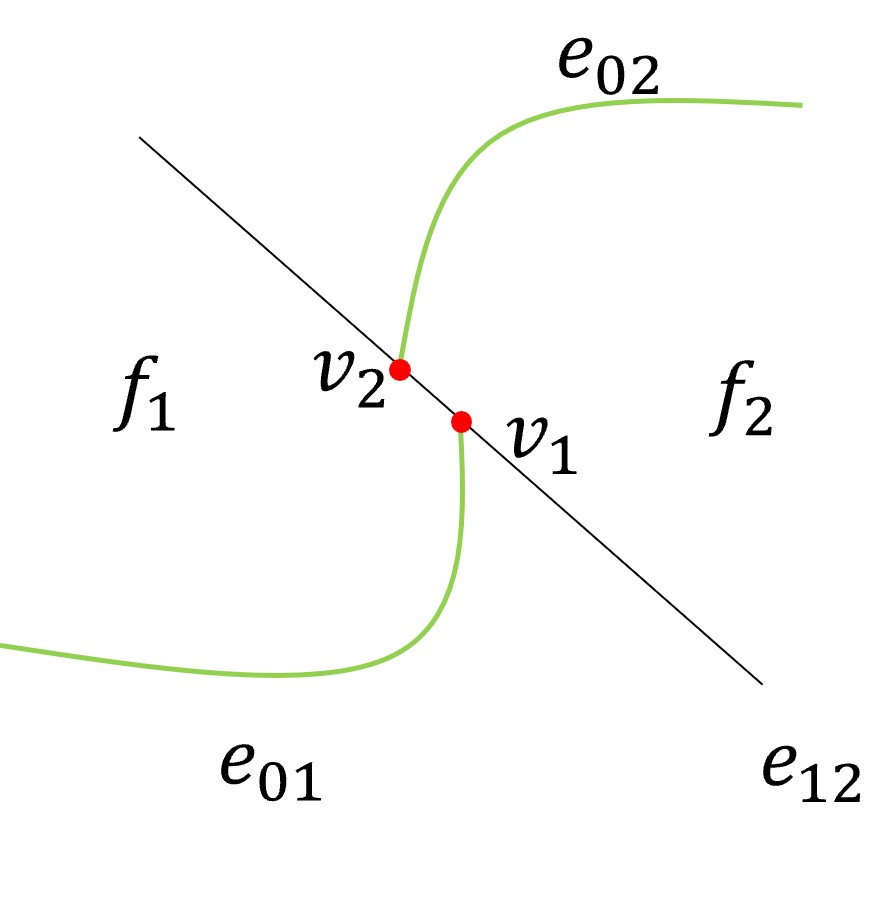}
    \caption{Sketch map of the disconnection of two intersection edges. }
    \label{fig:dislocation}
\end{figure}

When computing the intersection result of two models face by face, the geometric intersection algorithm does not use the information of neighbor faces, which may cause some disconnections on the boundary of faces, shown in Fig.\ref{fig:dislocation}. Because of the accumulating errors, two intersection edges on two separate faces, which should be connected, break up on the common boundary of the two faces.  $e_{01}$ is the intersection edge of face $f_0$(not showing in this map) and face $f_1$, while $e_{02}$ is the intersection edge of face $f_0$ and face $f_2$. $e_{12}$ is the common edge of $f_1$ and $f_2$. $v_1$ is the intersection vertex of $e_{1}$ and $e_{2}$, while $v_1$ is the intersection vertex of $e_{01}$ and $e_{2}$. As $f_0$ crosses through $e_{12}$, $v_1$ and $v_2$ should be at the same position in theory. However, because the intersection algorithm calculates the intersections of $f_0$ and $f_1$, $f_0$ and $f_2$ separately, the results of the two intersections can be inconsistent due to the accumulation of multiple kinds of errors. This inconsistency is manifested by the dislocation of $v_1$ and $v_2$, which makes two intersection edges on the same face $f_0$ disconnected. Usually, the inaccuracy of the shape of intersection edges will not make the following boolean steps fail. However, in this situation, not only will this disconnection make the resulting models inaccurate, but more seriously, it will make the area that should be bounded by intersection edges open, which will result in an illegal topological structure and make the following boolean operations crash. Therefore, this disconnection must be fixed to ensure the resulting model is legal and right.

There are three criteria we used to judge whether this situation occurs on two intersection edges $e_{01}$ and $e_{02}$, and these criteria should be satisfied simultaneously:

\noindent\ding{172} $e_{01}$ and $e_{02}$ have a pair of vertices $v_1$ and $v_2$ belonging separately to these two intersection edges that are very closed judged by the feature-based tolerance values of the edges.

\noindent\ding{173} $e_{01}$ and $e_{02}$ have a common face $f_0$ which generates them from the intersection between itself and other two separate faces $f_1$ and $f_2$.

\noindent\ding{174} $f_1$ and $f_2$ share common edges, and $v_1$ and $v_2$ are both on one of the common edges $e_{12}$.

The proof of why the three criteria satisfy this situation is shown as follows:

\setcounter{equation}{0}
\begin{proo}
from criterion \ding{173}, we know that:
\begin{align}
    &e_{01} = f_0\cap f_1\\
    &e_{02} = f_0\cap f_2
\end{align}
from criterion \ding{172} and criterion \ding{174}, we know that:
\begin{align}
    &v_{1} = e_{01}\cap e_{12}\\
    &v_{2} = e_{02}\cap e_{12}
\end{align}
from criterion \ding{174}, we know that:
\begin{align}
    &e_{12} = f_{1}\cap f_{2}
\end{align}
Therefore, from (3), (1), and (5), we know that:
\begin{align}
    v_1 &= e_{01}\cap e_{12}\notag\\
        &= (f_{0}\cap f_{1})\cap(f_1\cap f_2) \notag\\
        &= f_0\cap(f_1\cap f_2)\notag\\
        &= f_0\cap e_{12}
\end{align}
Similarly, we can get that
\begin{align}
    v_2=f_0\cap e_{12}
\end{align}
From criterion \ding{172}, we know that $v_1$ and $v_2$ are very close. Therefore, from (6) and (7), we know that $v_1$ and $v_2$ should be at the same position, and the position is the position of the intersection vertex of $f_0$ and $e_{12}$
\qed
\end{proo}

\begin{algorithm}[H]
    \caption{Judge Whether the Disconnection of Two Intersection Edges Occurs}
    \label{algorithm:judge1}
    \SetKw{True}{true}
    \SetKw{False}{false}
    \SetKw{Is}{is}
    \SetKw{And}{and}
    \SetKw{Break}{break}
    \SetKw{Maximum}{maximum}
    \SetKwFunction{GetStartAndEndVertex}{getStartAndEndVertex}
    \SetKwFunction{DistanceBetween}{distanceBetween}
    \SetKwFunction{GetFeatureBasedTolerance}{getFeatureBasedTolerance}
    \SetKwFunction{FindNearestPair}{findNearestPair}
    \SetKwFunction{GetSourceFaces}{getSourceFaces}
    \SetKwFunction{HaveCommonFace}{haveCommonFace}
    \SetKwFunction{GetCommonFace}{getCommonFace}
    \SetKwFunction{GetDifferentFace}{getDifferentFace}
    \SetKwFunction{FeatureValue}{featureValue}
    \SetKwFunction{GetPos}{getPos}
    \SetKwFunction{GetSharedEdges}{getSharedEdges}
    \SetKwFunction{PosOnEdge}{posOnEdge}
    \SetKwFunction{SetPos}{setPos}
    \SetKwFunction{EdgeFaceIntersection}{edgeFaceIntersection}
    \SetKwInOut{Input}{input}
    \SetKwInOut{Output}{output}
    \Input{$e_1$, $e_2$}
    \Output{Whether two intersection edges meet all three criteria}
    $v_{11},v_{12} \gets\GetStartAndEndVertex(e_1)$\;
    $v_{21},v_{22} \gets\GetStartAndEndVertex(e_1)$\;
    $t\gets\GetFeatureBasedTolerance(e_1) + \GetFeatureBasedTolerance(e_2)$\;
    $v_1, v_2\gets\FindNearestPair(\{v_{11}, v_{12}\}, \{v_{21}, v_{22}\})$\;
    \If{$\DistanceBetween(v_1, v_2) > t$}{
        \Return \False\;
    }
    $f_{11}, f_{12}\gets\GetSourceFaces(e_1)$\;
    $f_{21}, f_{22}\gets\GetSourceFaces(e_2)$\;
    \If{$\HaveCommonFace(\{f_{11}, f_{12}\},\{f_{21}, f_{22}\})$ \Is\False}{
        \Return \False\;
    }
    $f_0\gets\GetCommonFace(\{f_{11}, f_{12}\},\{f_{21}, f_{22}\})$\;
    $f_1, f_2\gets\GetDifferentFace(\{f_{11}, f_{12}\},\{f_{21}, f_{22}\})$\;
    $e\_array\gets\GetSharedEdges(f_1, f_2)$\;
    $flag\gets$\False\;
    $pos_1\gets\GetPos(v_1)$\;
    $pos_2\gets\GetPos(v_2)$\;
    \For{$edge\in e\_array$}{
        \If{$\PosOnEdge(pos_1, edge)$ \And $\PosOnEdge(pos_2, edge)$}{
            $flag\gets$\True\;
            $e_0\gets edge$\;
            \Break\;
        }
    }
    \If{$flag$ \Is\False}{
        \Return \False\;
    }
    \Return \True\;
\end{algorithm}

Therefore, when all three above conditions meet, it means this disconnection situation exists. $v_1$ and $v_2$ should be at the same position. Another crucial thing from the proof is that a proper position to merge these two misplaced vertices is the position of the intersection vertex generated by the intersection of the common edge $e_{12}$ and face $f_0$. The algorithm that detects this situation is shown as \textbf{Algorithm \ref{algorithm:judge1}}.

Since the dimension of the edge-face intersection is lower than that of the face-face intersection, which means lower errors, the new position will be more precise than the old one. The repair process is shown as \textbf{Algorithm \ref{algorithm:fix1}}.

\begin{algorithm}[H]
    \caption{Fix the Disconnection of two Intersection Edges}
    \label{algorithm:fix1}
    \SetKw{True}{true}
    \SetKw{False}{false}
    \SetKw{Is}{is}
    \SetKw{And}{and}
    \SetKw{Break}{break}
    \SetKw{Maximum}{maximum}
    \SetKwFunction{GetStartAndEndVertex}{getStartAndEndVertex}
    \SetKwFunction{DistanceBetween}{distanceBetween}
    \SetKwFunction{GetFeatureBasedTolerance}{getFeatureBasedTolerance}
    \SetKwFunction{FindNearestPair}{findNearestPair}
    \SetKwFunction{GetSourceFaces}{getSourceFaces}
    \SetKwFunction{HaveCommonFace}{haveCommonFace}
    \SetKwFunction{GetCommonFace}{getCommonFace}
    \SetKwFunction{GetDifferentFace}{getDifferentFace}
    \SetKwFunction{FeatureValue}{featureValue}
    \SetKwFunction{GetPos}{getPos}
    \SetKwFunction{GetSharedEdges}{getSharedEdges}
    \SetKwFunction{OnEdge}{onEdge}
    \SetKwFunction{SetPos}{setPos}
    \SetKwFunction{EdgeFaceIntersection}{edgeFaceIntersection}
    \SetKwInOut{Input}{input}
    \Input{$e_1$, $e_2$}
    $v\_array\gets\EdgeFaceIntersection(e_0, f_0)$\;
    $min\_dist\gets\Maximum$\;
    \For{$v\in v\_array$}{
        $dist\gets \DistanceBetween(v,v_1) + \DistanceBetween(v,v_2)$\;
        \If{$dist<min\_dist$}{
            $pos\gets\GetPos(v)$\;
            $min\_dist\gets dist$\;
        }
    }
    $\SetPos(v_1, pos)$\;
    $\SetPos(v_2, pos)$\;
    \Return\;
\end{algorithm}

\subsubsection{Pretty Short Edges which Should Not Exist}
\label{subsubsection:Pretty Short Edges}

It is not rare that the intersection result of two models has some very short edges. Their length is pretty small compared with the size of the models, and we should deal with them very carefully. Sometimes a pretty short edge is the remnant of a normal-length curve cut by faces' borders. But in some situations, this short edge comes from the intersection result of two slightly contacting faces, like two faces sharing a common vertex, or two tangent faces. And when these situations occur, the intersection edge should collapse into an intersection vertex based on the edge-face topological intersection results. However, the errors may disturb the process of topological edge-face intersection and make a small part of the intersection edge left.  

\begin{figure}[H]
    \centering
    \includegraphics[scale=0.8]{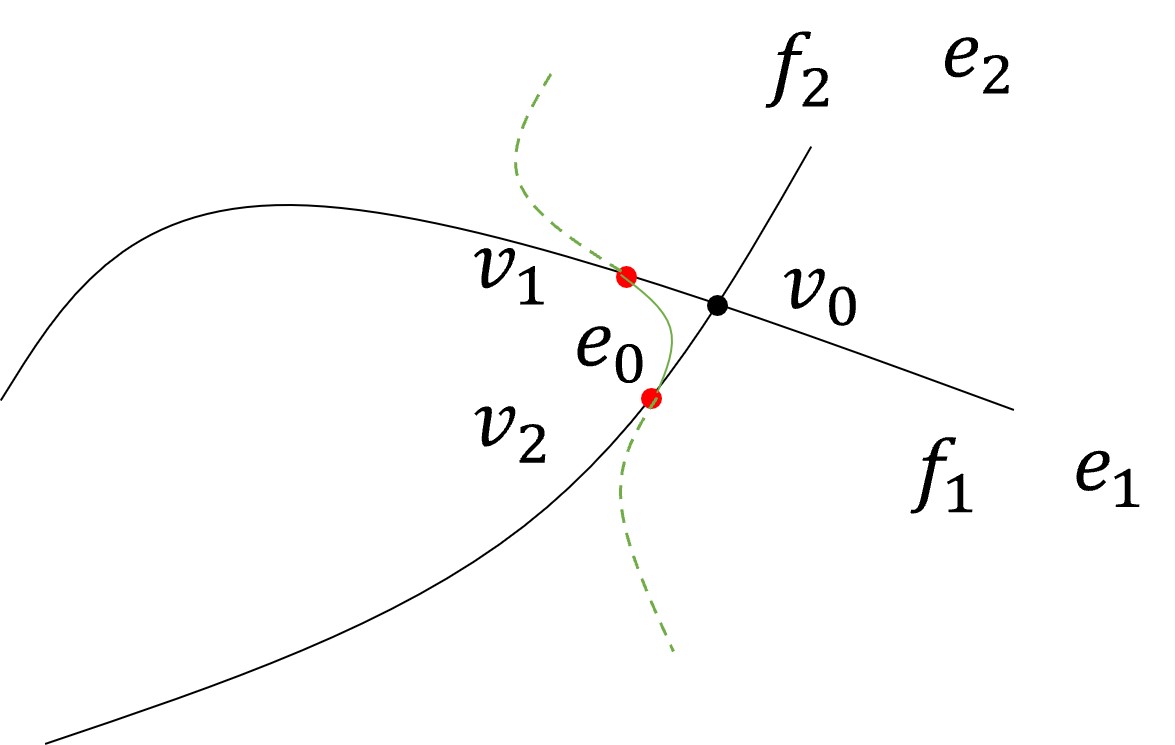}
    \caption{Sketch map of the pretty short edge.}
    \label{fig:situation2}
\end{figure}

When two faces only intersect at certain positions, their overlapping areas are so small that the influence of accumulating errors will be obvious. As shown in Fig.\ref{fig:situation2},  $e_1$ is a boundary edge of face $f_1$,  while $e_2$ is a boundary edge of face $f_2$. $e_0$ is the valid part of intersection curve of $f_1$ and $f_2$(The invalid parts which are outside $f_1$ or $f_2$ are shown by dashed lines). This intersection curve should pass to the $v_0$, the intersection vertex of $e_1$ and $e_2$. But due to accumulating errors, the intersection curve intersects $e_1$ at $v_1$ and intersects $e_2$ at $v_2$, which causes a pretty short intersection edge $e_0$. And the pretty short edges will bind a sliver face and falsely break up the boundary edge of faces, which will cause topological problems in the resulting model.

We have four criteria to judge whether a pretty short edge $e_0$ in an intersection graph $g$ should be an intersection vertex and need to be fixed, and these criteria should be satisfied in full:

\noindent\ding{172}$e_0$ is an intersection edge of face $f_1$ and $f_2$ while the length of $e_0$ is shorter than its feature based tolerance.

\noindent\ding{173} $e_0$ in $g$ does not connect with any other edges or only connects to other edges in one end.

\noindent\ding{174} $e_1$ is a boundary edge of $f_1$. $e_2$ is a boundary edge of $f_2$. $e_0$ intersects with $e_1$ at the position of $v_1$. $e_0$ intersects with $e_2$ at the position of $v_2$.

\noindent\ding{175} $e_1$ intersects with $e_2$ at the position of $v_0$. Both the distances between $v_0$ and $v_1$, $v_0$, and $v_2$ are smaller than the sum of feature-based tolerance of $e_1$ and $e_2$.

The proof of why the four criteria satisfy this situation is shown as follows:

\setcounter{equation}{0}
\begin{proo}
from criterion \ding{172}, we know that:
\begin{align}
    &e_{0} = f_1\cap f_2
\end{align}
from criterion \ding{174}, we know that:
\begin{align*}
    &e_{1}\in f_1\\
    &e_{2}\in f_2
\end{align*}
Therefore, based on \textbf{Rule \ref{rule1}}, we know that
\begin{align}
    &e_{1} = e_{1}\cap f_{1}\\
    &e_{2} = e_{2}\cap f_{2}
\end{align}
from criterion \ding{175} and (2), we know that
\begin{align}
 v_0 &= e_1\cap e_2\notag\\
     &= e_1\cap f_1\cap e_2\notag\\
     &= e_1\cap (f_1\cap e_2)
\end{align}
Therefore, from (4), we know that
\begin{align}
    v_0\in f_1\cap e_2
\end{align}
Based on (5) and \textbf{Lemma \ref{lemma1}}, in a sufficiently small neighborhood of $v_0$, we can get that
\begin{align}
    v_0 = f_1\cap e_2
\end{align}
from criterion \ding{174}, we know that
\begin{align}
    v_2&=e_2\cap e_0
\end{align}
Then, from (7), (1), and (2), we know that:
\begin{align}
    v_2&=(e_2\cap f_2)\cap (f_1\cap f_2)\notag\\
       &= e_2\cap f_1 \cap f_2\notag\\
       &= f_2\cap (f_1\cap e_2)
\end{align}
Therefore, from (8), we know that
\begin{align}
    v_2\in f_1\cap e_2
\end{align}
Based on (8) and \textbf{Lemma \ref{lemma1}}, in a sufficiently small neighborhood of $v_2$, we can get that
\begin{align}
    v_2 = f_1\cap e_2
\end{align}
From criterion \ding{175}, we know that $v_0$ and $v_2$ are very close. Therefore, based on (7) and (10), we can get that
\begin{align}
    v_2 =v_0 = f_1\cap e_2
\end{align}
Similarly, we can get that 
\begin{align}
    v_1 =v_0
\end{align}

From criterion \ding{173}, we know that removing $e_0$ from the intersection graph will not break the graph apart. From (10), (11) and the information that $e_0$ is very short, we know that the intersection edge should collapse into a vertex at the same position of $v_0$. 
\qed
\end{proo}

Length is an important criterion, but we need more than just considering this factor to identify flawed edges. In our practice, we find some edges very short but valid. These edges have such a feature that they are at a crucial position in the intersection graph. If we remove them, the intersection graph will be split up, and the areas that intersection edges should bound will be open, which will cause an incomplete topological structure in the resulting model. Therefore, the second criterion can avoid this mistake. Besides, if this situation happens totally inside a face, it will not affect any existing topological structure. It will add an isolated loop inside the face, which will be easily detected and removed in the final resulting model, so we do not need to do extra work here to fix it. That is why we need the third criterion. And the fourth condition restricts all steps to a small area. We repair the short edge only when the intersection edge meets all four requirements. The judge algorithm is shown as \textbf{Algorithm \ref{algorithm: judge2}}.

\begin{algorithm}[H]
    \caption{Judge Whether the Short Edge Should Exist}
    \label{algorithm: judge2}
    \SetKw{True}{true}
    \SetKw{False}{false}
    \SetKw{Is}{is}
    \SetKw{Not}{not}
    \SetKw{And}{and}
    \SetKw{Or}{or}
    \SetKw{Break}{break}
    \SetKw{Exist}{exist}
    \SetKw{Maximum}{maximum}
    \SetKw{Null}{null}
    \SetKwFunction{GetStartAndEndVertex}{getStartAndEndVertex}
    \SetKwFunction{DistanceBetween}{distanceBetween}
    \SetKwFunction{GetFeatureBasedTolerance}{getFeatureBasedTolerance}
    \SetKwFunction{FindNearestPair}{findNearestPair}
    \SetKwFunction{GetSourceFaces}{getSourceFaces}
    \SetKwFunction{HaveCommonFace}{haveCommonFace}
    \SetKwFunction{GetCommonFace}{getCommonFace}
    \SetKwFunction{GetDifferentFace}{getDifferentFace}
    \SetKwFunction{FeatureValue}{featureValue}
    \SetKwFunction{GetPos}{getPos}
    \SetKwFunction{GetSharedEdges}{getSharedEdges}
    \SetKwFunction{PosOnEdge}{PosOnEdge}
    \SetKwFunction{SetPos}{setPos}
    \SetKwFunction{EdgeFaceIntersection}{edgeFaceIntersection}
    \SetKwFunction{GetLength}{getLength}
    \SetKwFunction{GetNext}{getNext}
    \SetKwFunction{GetPrevious}{getPrevious}
    \SetKwFunction{PosOnBoundary}{posOnBoundary}
    \SetKwFunction{GetPosOnFaceBoundary}{GetposOnFaceBoundary}
    \SetKwFunction{EdgeEdgeIntersection}{edgeEdgeIntersection}
    \SetKwInOut{Input}{input}
    \SetKwInOut{Output}{output}
    \Input{intersection edge $e_0$, intersection graph $g$}
    \Output{Whether this edge needs to be repaired. If so, return $v_0$, whose position is that the pretty short edges should collapse into.}
    $t\gets\GetFeatureBasedTolerance(e_0)$\;
    \If{$\GetLength(e_0) > t$}{
        \Return \False\;
    }
    \If{$\GetNext(g, e_0)$ \Is\Exist\And $\GetPrevious(g, e_0)$ \Is\Exist}{
        \Return \False\;
    }
    $v_{1},v_{2} \gets\GetStartAndEndVertex(e_0)$\;
    $f_{1},f_{2}\gets\GetSourceFaces(e_0)$\;
    $pos_1\gets\GetPos(v_1)$\;
    $pos_2\gets\GetPos(v_2)$\;
    \If{$\PosOnBoundary(pos_1, f_1)$\Is\True}{
        $e_1\gets\GetPosOnFaceBoundary(pos_1, f_1)$\;
    }    
    \If{$\PosOnBoundary(pos_1, f_2)$\Is\True}{
        $e_2\gets\GetPosOnFaceBoundary(pos_1, f_2)$\;
    }
    \If{$\PosOnBoundary(pos_2, f_1)$\Is\True}{
        $e_1\gets\GetPosOnFaceBoundary(pos_2, f_1)$\;
    }
    \If{$\PosOnBoundary(pos_2, f_2)$\Is\True}{
        $e_2\gets\GetPosOnFaceBoundary(pos_2, f_2)$\;
    }
    \If{$e_1$ \Is\Not\Exist\Or $e_2$ \Is\Not\Exist}{
        \Return \False\;
    }
    $v\_array\gets\EdgeEdgeIntersection(e_1, e_2)$\;
    $sum\gets\GetFeatureBasedTolerance(e_1) + \GetFeatureBasedTolerance(e_2)$\;
    \For{$v\in v\_array$}{
        $dist_1\gets \DistanceBetween(v_1, v)$\;
        $dist_2\gets \DistanceBetween(v_2, v)$\;
        \If{$dist_1<sum$ \And $dist_2<sum$}{
            $v_0\gets v$\;
            \Return \True, $v_0$\;
        }
    }
    \Return \False\;
\end{algorithm}

To fix this wrong edge, we should make these pretty short edges collapse into an isolated vertex, so it will no longer harm the existing topological structure. Moreover, we should find a proper position for this isolated vertex. In \textbf{Algorithm \ref{algorithm: judge2}}, we have found the position should be the position of $v_0$. Therefore, we should change the position of $v_1$ and $v_2$ into the position of $v_0$. Then we should merge $v_1$ and $v_2$ into one single vertex and remove the geometric element of $e_0$ since $e_0$ has already collapsed into a vertex. The process is shown as \textbf{Algorithm \ref{algorithm:fix2}}.

\begin{algorithm}[H]
    \caption{Fix the Pretty Short Edge by Collapsing It into Vertex}
    \label{algorithm:fix2}
    \SetKw{True}{true}
    \SetKw{False}{false}
    \SetKw{Is}{is}
    \SetKw{And}{and}
    \SetKw{Or}{or}
    \SetKw{Break}{break}
    \SetKw{Exist}{exist}
    \SetKw{Maximum}{maximum}
    \SetKw{Null}{null}
    \SetKwFunction{GetStartAndEndVertex}{getStartAndEndVertex}
    \SetKwFunction{DistanceBetween}{distanceBetween}
    \SetKwFunction{GetFeatureBasedTolerance}{getFeatureBasedTolerance}
    \SetKwFunction{FindNearestPair}{findNearestPair}
    \SetKwFunction{GetSourceFaces}{getSourceFaces}
    \SetKwFunction{HaveCommonFace}{haveCommonFace}
    \SetKwFunction{GetCommonFace}{getCommonFace}
    \SetKwFunction{GetDifferentFace}{getDifferentFace}
    \SetKwFunction{FeatureValue}{featureValue}
    \SetKwFunction{MergeVertex}{mergeVertex}
    \SetKwFunction{GetPos}{getPos}
    \SetKwFunction{GetSharedEdges}{getSharedEdges}
    \SetKwFunction{PosOnEdge}{PosOnEdge}
    \SetKwFunction{SetPos}{setPos}
    \SetKwFunction{EdgeEdgeIntersection}{edgeEdgeIntersection}
    \SetKwFunction{GetLength}{getLength}
    \SetKwFunction{GetNext}{getNext}
    \SetKwFunction{GetPrevious}{getPrevious}
    \SetKwFunction{PosOnBoundary}{posOnBoundary}
    \SetKwFunction{GetCurve}{getCurve}
    \SetKwFunction{SetCurve}{setCurve}
    \SetKwFunction{GetParam}{getParam}
    \SetKwFunction{eval}{eval}
    \SetKwInOut{Input}{input}
    \SetKwInOut{Output}{output}
    \Input{intersection edge $e_0$, intersection vertex $v_0$}
    $v_{1},v_{2} \gets\GetStartAndEndVertex(e_0)$\;
    $pos_0\gets\GetPos(v_0)$\;
    $\SetPos(v_1, pos_0)$\;
    $\SetPos(v_2, pos_0)$\;
    $\MergeVertex(v_1, v_2)$\;
    $\SetCurve(e_0, \Null)$\;
\end{algorithm}

\subsubsection{Deviation of Intersection Vertices}
\label{subsubsection:Inconsistency of Vertices}

In the process of face-face intersection, the initial intersection edge of these two faces will be constructed first based on the geometrical intersection curve. Then this edge will intersect with each boundary edge of the two faces to find valid parts inside both faces. However, as the process of edge-edge intersection runs edge by edge separately, some inconsistency between two neighbor edges may exist. 

\begin{figure}[H]
    \centering
    \includegraphics[scale=0.9]{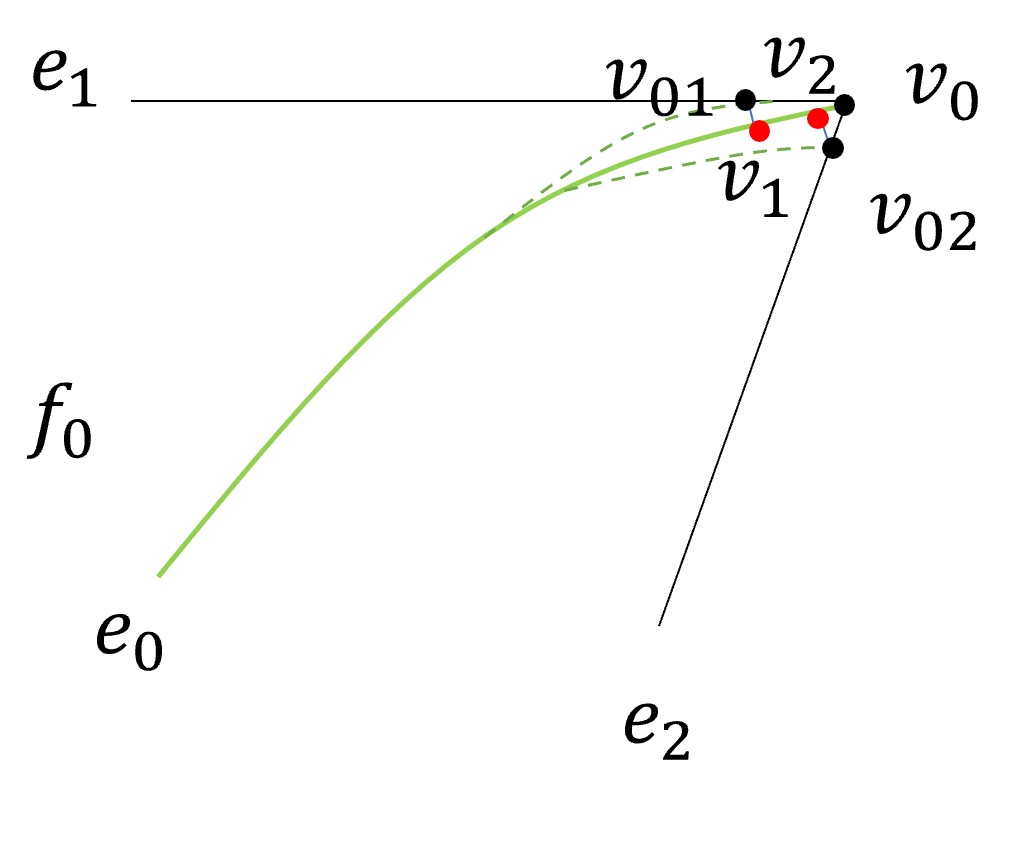}
    \caption{Sketch map of the deviation of intersection vertices}
    \label{fig:situation3}
\end{figure}

Shown in Fig.\ref{fig:situation3}, $e_0$ is the initial intersection edge of $f_0$ and $f_1$(not showing in this figure), $e_1$ and $e_2$ are two boundary edges of $f_0$, and they meet at a common vertex $v_0$. In theory, $e_0$ should have intersected with $e_1$ and $e_2$ at $v_0$. However, due to the calculation errors, when we intersect $e_0$ with $e_1$ and $e_2$ separately, we find that $e_0$ intersects with $e_1$ at $v_1$, and intersects with $e_2$ at $v_2$. This slight difference will cut $e_0$ and generate a short edge bounded by $v_1$ and $v_2$, which may damage the topological integrity.

When dealing with $e_0$ intersects with $e_1$, the geometric intersection algorithm gives the intersection position of $v_{01}$ on $e_1$. Use the position of $v_{01}$ to find the corresponding parameter inversely on $e_0$. However, when applying the parameter on $e_0$, we get the position of $v_1$. The position of $v_1$ and $v_{01}$ is slightly different due to errors. The topological algorithm uses the parameter to get the position of the intersection vertex. Therefore, $v_1$ is the topological intersection result. And similar things happen when dealing with $e_0$ intersects with $e_2$, which generates $v_2$ and $v_{02}$. And $v_2$ is the topological intersection result. This situation is like the disconnection of two intersection edges we have discussed in \textbf{section \ref{subsubsection:Disconnection}}. However, unlike happening at a common edge of two neighbor faces,  it exists on two neighbor edges sharing a common vertex.

To detect this situation, we have four criteria to judge whether an intersection edge $e_0$ needs to be repaired, and these criteria should be satisfied in full:

\noindent\ding{172} $e_0$ generated by the intersection of $f_0$ and $f_1$ intersects with at least two boundary edges of $f_0$.

\noindent\ding{173} $e_1$ and $e_2$ are boundary edges of $f_0$. $e_0$ on $f_0$ intersects with $e_1$ at $v_1$, $e_2$ at $v_2$, while $e_1$ and $e_2$ share common vertex $v_0$.

\noindent\ding{174} The distances between $v_1$ and $v_0$, $v_2$ and $v_0$ are both shorter than the feature based tolerance of $e_0$.

\noindent\ding{175} $v_0$ is on $f_1$.

The proof of why the four criteria satisfy this situation is shown as follows:

\setcounter{equation}{0}
\begin{proo}
from criterion \ding{172}, we know that:
\begin{align}
    &e_{0} = f_0\cap f_1
\end{align}
from criterion \ding{173}, we know that:
\begin{align*}
    &e_{1}\in f_0\\
    &e_{2}\in f_0
\end{align*}
Therefore, based on \textbf{Rule \ref{rule1}}, we know that
\begin{align}
    &e_{1} = e_{1}\cap f_{0}\\
    &e_{2} = e_{2}\cap f_{0}
\end{align}
from criterion \ding{173}, we know that
\begin{align}
    e_0\cap e_1 = v_1\\
    e_0\cap e_2 = v_2
\end{align}
Therefore, from (4), (2) and (1), we know that
\begin{align}
    v_1 &= e_0\cap e_1\notag\\
        &= (f_0\cap f_1)\cap e_1\notag\\
        &= f_1\cap (f_0\cap e_1)\notag\\
        &= f_1\cap e_1
\end{align}
from criterion \ding{173}, we know that
\begin{align}
    v_0 = e_1\cap e_2
\end{align}
from criterion \ding{175} and (7), we know that
\begin{align}
    v_0 = e_1\cap e_2\cap f_1\notag
\end{align}
Therefore, we know that
\begin{align}
    v_0 \in e_1\cap f_1
\end{align}
Based on (8) and \textbf{Lemma \ref{lemma1}}, in a sufficiently small neighborhood of $v_0$, we can get that
\begin{align}
    v_0 = e_1\cap f_1
\end{align}
From criterion \ding{174}, we know that $v_0$ and $v_1$ are very close, therefore, based on (6) and (9), we can get that
\begin{align}
    v_1 =v_0 = e_1\cap f_1
\end{align}
Similarly, we can get that
\begin{align}
    v_2 =v_0
\end{align}
Therefore, $v_1$ and $v_2$ should be removed and we use $v_0$ to replace them.
\qed
\end{proo}

Because we have found that $v_0$ is on $f_1$, and $v_0$ originally belongs to $f_0$, we can assert that $v_0$ must be on the intersection edge of $f_0$ and $f_1$. Therefore, the process to fix $e_0$ is simple. We just need to remove $v_1$ and $v_2$ and then use $v_0$ to substitute them. 

Unlike other scenes, in this situation, the flawed intersection edge should be fixed during the process of face-face intersection. Otherwise, some intersection information that can help fix the inconsistency will be lost. In the face-face intersection, $e_0$, the initial intersection edge of $f_0$ and $f_1$ will be used to intersect with all boundary edges of $f_0$ and $f_1$ to find its valid parts, which are both inside $f_0$ and $f_1$. If $e_0$ intersects with $e_1$ at the position of $v_1$ and intersects with $e_2$ at the position of $v_2$, while $e_1$ and $e_2$ belong to the same face, for example, $f_0$. Then it has fit the criterion \ding{172}, and the algorithm shown in \textbf{Algorithm \ref{algorithm:Judge and fix the inconsistency of vertices}} will be called.

\begin{algorithm}[H]
    \label{algorithm:Judge and fix the inconsistency of vertices}
    \caption{Judge and Fix the Deviation of Intersection Vertices on an Intersection Edge}
    \SetKw{True}{true}
    \SetKw{False}{false}
    \SetKw{Is}{is}
    \SetKw{Not}{not}
    \SetKw{And}{and}
    \SetKw{Or}{or}
    \SetKw{Break}{break}
    \SetKw{Exist}{exist}
    \SetKw{Maximum}{maximum}
    \SetKw{Null}{null}
    \SetKwFunction{GetStartAndEndVertex}{getStartAndEndVertex}
    \SetKwFunction{DistanceBetween}{distanceBetween}
    \SetKwFunction{GetFeatureBasedTolerance}{getFeatureBasedTolerance}
    \SetKwFunction{FindNearestPair}{findNearestPair}
    \SetKwFunction{GetSourceFaces}{getSourceFaces}
    \SetKwFunction{HaveCommonFace}{haveCommonFace}
    \SetKwFunction{GetCommonFace}{getCommonFace}
    \SetKwFunction{GetCommonVertex}{getCommonVertex}
    \SetKwFunction{GetPosOnFaceBoundary}{GetPosOnFaceBoundary}
    \SetKwFunction{GetDifferentFace}{getDifferentFace}
    \SetKwFunction{FeatureValue}{featureValue}
    \SetKwFunction{GetPos}{getPos}
    \SetKwFunction{GetSharedEdges}{getSharedEdges}
    \SetKwFunction{PosOnEdge}{PosOnEdge}
    \SetKwFunction{SetPos}{setPos}
    \SetKwFunction{EdgeEdgeIntersection}{edgeEdgeIntersection}
    \SetKwFunction{GetLength}{getLength}
    \SetKwFunction{GetNext}{getNext}
    \SetKwFunction{GetPrevious}{getPrevious}
    \SetKwFunction{PosOnBoundary}{posOnBoundary}
    \SetKwFunction{PosOnFace}{posOnFace}
    \SetKwFunction{Remove}{remove}
    \SetKwFunction{Add}{add}
    \SetKwFunction{GetCurve}{getCurve}
    \SetKwFunction{SetCurve}{setCurve}
    \SetKwFunction{GetParam}{getParam}
    \SetKwFunction{eval}{eval}
    \SetKwInOut{Input}{input}
    \SetKwInOut{Output}{output}
    \Input{intersection edge $e_0$, $f_0$ and $f_1$ which generate intersection edge $e_0$, vertex $v_1$ and $v_2$ generated by the intersection between $e_0$ and the boundary edge of $f_0$.}
    $pos_1\gets\GetPos(v_1)$\;
    $pos_2\gets\GetPos(v_2)$\;
    $t\gets\GetFeatureBasedTolerance(e_0)$\;
    $e_1\gets\GetPosOnFaceBoundary(pos_1, f_0)$\;
    $e_2\gets\GetPosOnFaceBoundary(pos_2, f_0)$\;
    \If{$\GetCommonVertex(e_1, e_2)\ \Is\ \Not\ \Exist$}{
        \Return\False\;
    }
    \For{$v\in \GetCommonVertex(e_1, e_2)$}{
        $dist_1\gets \DistanceBetween(v_1, v)$\;
        $dist_2\gets \DistanceBetween(v_2, v)$\;
        \If{$dist_1<t$ \And$dist_2<t$}{
            $v_0\gets v$\;
            $pos_0\gets\GetPos(v_0)$\;
            \If{$\PosOnFace(pos, f_1)$\Is\True}{
                $\Remove(v_{array}, v_1)$\;
                $\Remove(v_{array}, v_2)$\;
                $\Add(v_{array}, v_0)$\;
                \Return\True\;
            }
        }
    }
    \Return\False\;
\end{algorithm}

\section{Experiment}
In this part, we will show our repair method in different situations. Our method can detect and fix flawed structures while keeping the flawless parts of models from being modified. 
\begin{example}
As shown in Fig.\ref{fig:exp1}, the torus's center position is the same as the center position of the regular hexagonal prism, and the torus's axis of rotation coincides with the center axis of the regular hexagonal prism. The minor diameter of the torus, which is the diameter of the circle being rotated, is the same as the height of the prism. The major radius of the torus, which is the distance from the center of the circle to its rotation axis, is the same as the radius of the circumscribed circle of the bottom plane of the prism. Therefore, all vertices of the prism are on the face of the torus. And there should be 12 intersection edges between these two bodies.

\begin{figure}[H]
    \centering
    \subcaptionbox{}{
        \includegraphics[scale=0.55]{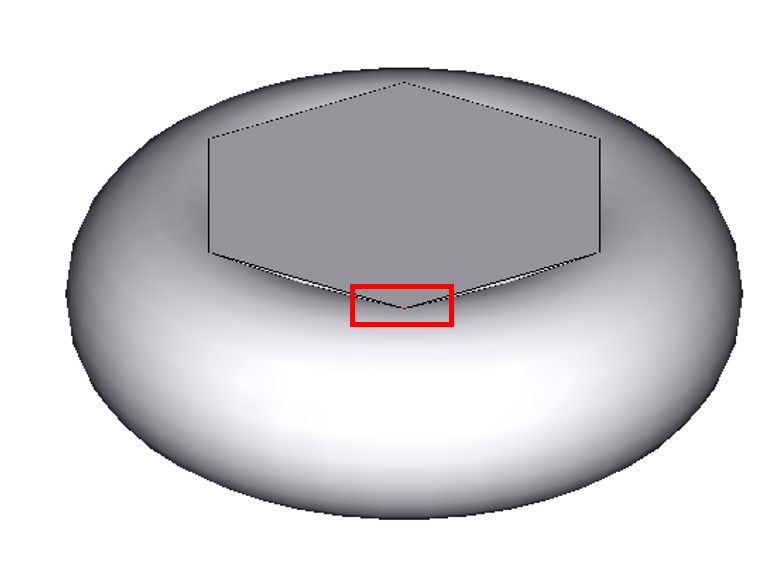}
    }
    \subcaptionbox{}{
        \includegraphics[scale=0.5]{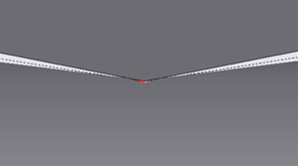}
    }
    \caption{(a) Relative position of the regular hexagonal prism and the torus. (b) 
An abnormal short edge appears in the boolean union resulting model of the prism and the torus.}
    \label{fig:exp1}
\end{figure}

However, in practice, the number of intersection edges is more than 12 because of the intersection errors. And there are some pretty short edges appearing near the position of the prism's vertices. This situation is described in \textbf{section \ref{subsubsection:Inconsistency of Vertices}}. Fortunately, these short edges will not fail the following boolean operations, but they are still unexpected and will make the resulting model's structures more complex, which may cause potential problems.  

Using the algorithm described in \textbf{section \ref{subsubsection:Inconsistency of Vertices}}, we can easily detect the short edges and prolong its neighbor intersection edge to meet the vertex of the prism. After repairing, all tiny intersection edges are removed. And so will the short edges left in the resulting model.  The boolean result will be more concise than that before.

Although there are some pretty short intersection edges, it should be noted that this case is different from the situation described in \textbf{section \ref{subsubsection:Pretty Short Edges}}. These short edges, which are generated by the inconsistency of intersections between intersection edges and boundary edges, are connected in the intersection graph at both ends.

\end{example}

\begin{example}
As shown in Fig.\ref{fig:torus and prism}(a), the prism and torus change their relative position. in this situation, the prism translates a certain distance horizontally outward from the torus. Fig.\ref{fig:torus and prism}(b) shows these two bodies' intersection graph. However, this intersection graph is split into three parts for the reason discussed in \textbf{section \ref{subsubsection:Disconnection}}. 

\begin{figure}[H]
    \centering
    \subcaptionbox{}{
        \includegraphics[scale=0.45]{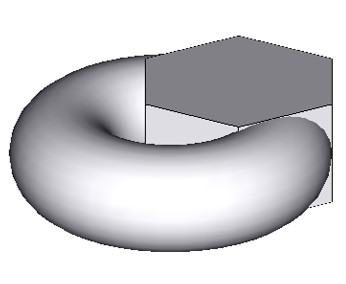}
    }
    \subcaptionbox{}{
        \includegraphics[scale=0.45]{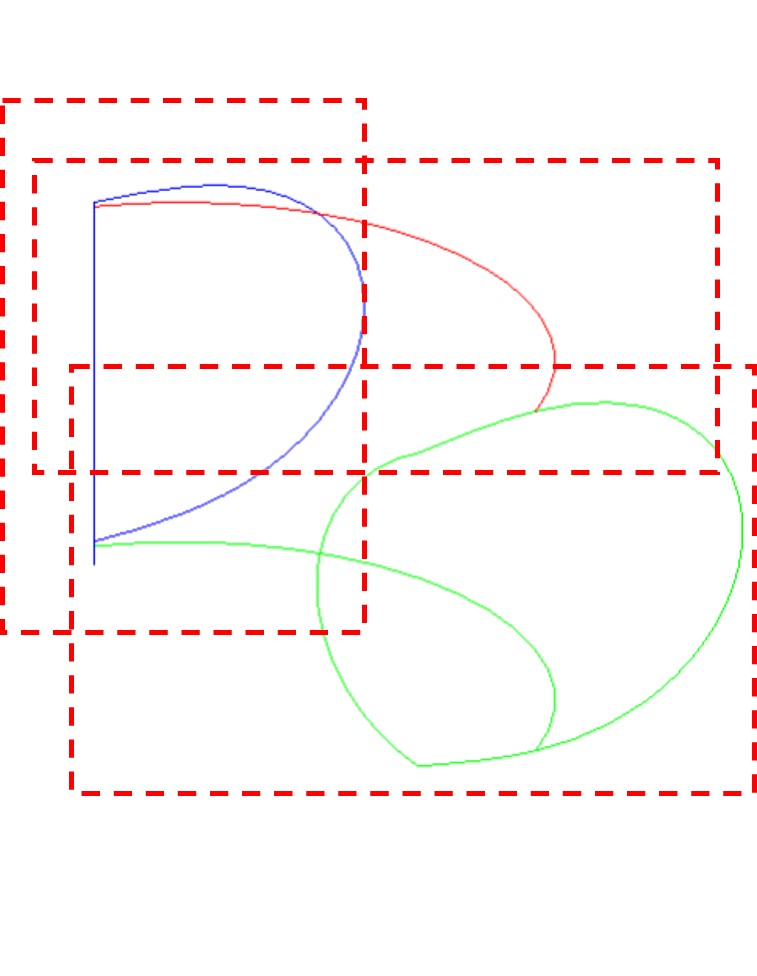}
    }
    \caption{(a) Relative position of the prism and torus. (b) Intersection graph of the prism and torus. Because of the disconnection, the graph has been split into three parts. Each part in different dashed boxes is dyed different colors.}
    \label{fig:torus and prism}
\end{figure}

As shown in Fig.\ref{fig:exp2-3}, intersection edges that should have shared one of their vertices break apart. Theoretically, these two vertices should be combined to make the intersection graph valid. And our inference procedure can find this flawed structure and fix it.

\begin{figure}[H]
    \centering
    \includegraphics[scale=0.45]{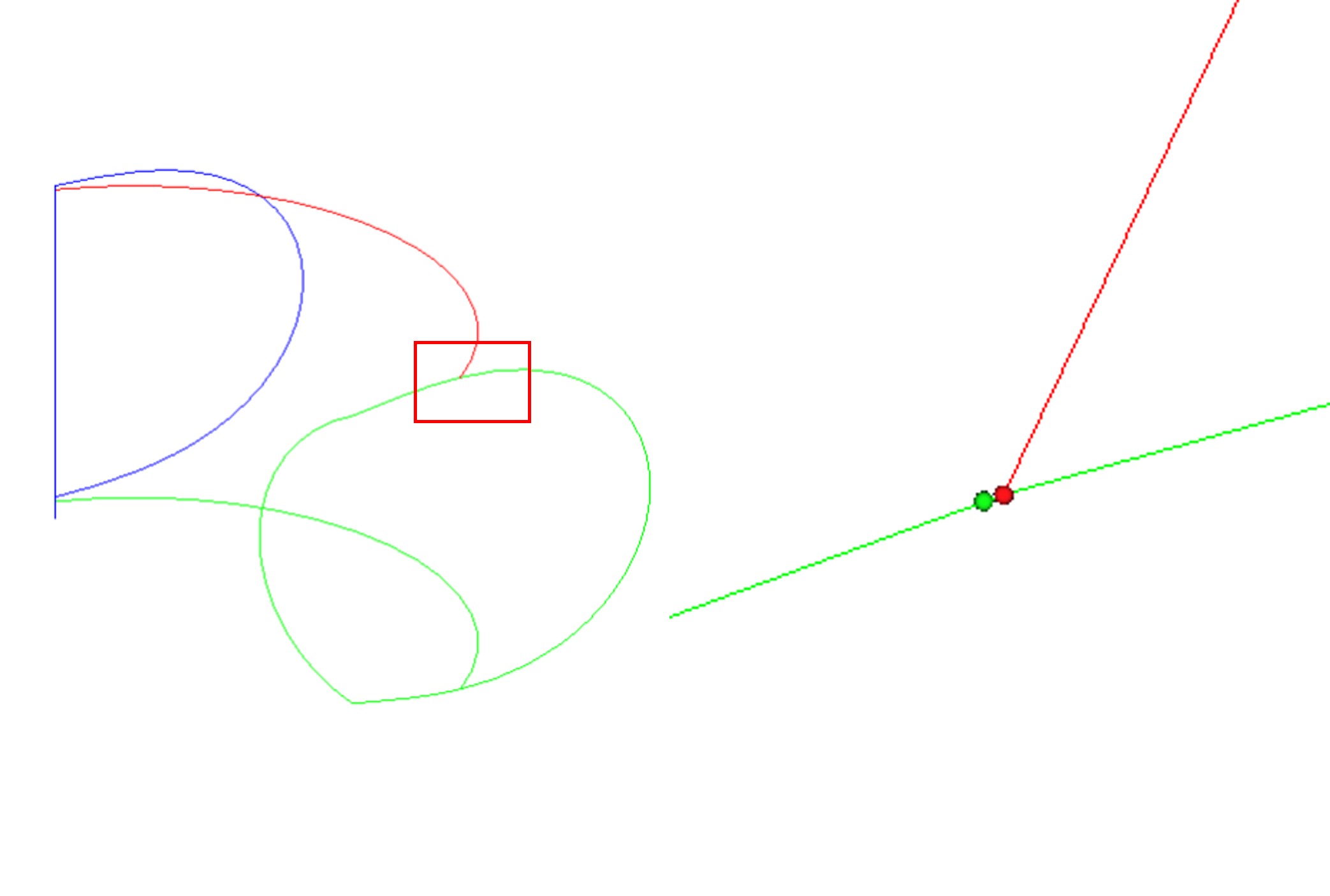}
    \caption{A pair of disconnected vertices in the intersection graph}
    \label{fig:exp2-3}
\end{figure}

Without repairing, the area that intersection edges should have bounded is open now. In the following steps, after imprinting intersection edges on these two bodies, there will be some faces falsely being segmented, shown in Fig.\ref{fig:torus and prism imprint result}. There should be three faces on the surface of the torus after imprinting, shown in Fig.\ref{fig:torus and prism imprint result} (a). However, due to the disconnection, one area which should be bounded by intersection edges is open now. Therefore, the false result only has two faces after imprinting, shown in Fig.\ref{fig:torus and prism imprint result} (b). Consequently, this will affect the following classification of faces, making some faces deleted by mistake and generating incorrect boolean operations, as shown in Fig.\ref{fig:torus and prism boolean result} (a).

\begin{figure}[H]
    \centering
    \subcaptionbox{}{
        \includegraphics[scale=0.3]{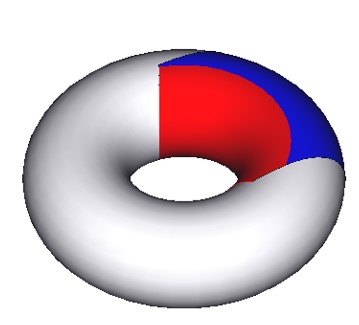}
    }
    \subcaptionbox{}{
        \includegraphics[scale=0.3]{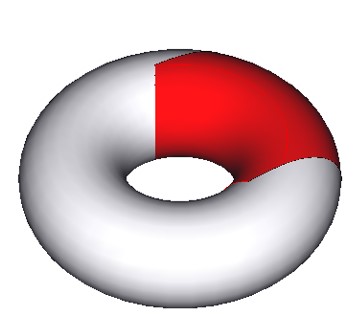}
    }
    \caption{(a) Right imprint result. (b) False imprint result due to the disconnection.}
    \label{fig:torus and prism imprint result}
\end{figure}

By using the algorithm discussed in \textbf{section \ref{subsubsection:Disconnection}}, we can reunite the disconnecting vertex. After repairing, the split intersection graph reunites, and boolean results are right now, shown in Fig.\ref{fig:torus and prism boolean result} (b).

\begin{figure}[H]
    \centering
    \subcaptionbox{}{
        \includegraphics[scale=0.3]{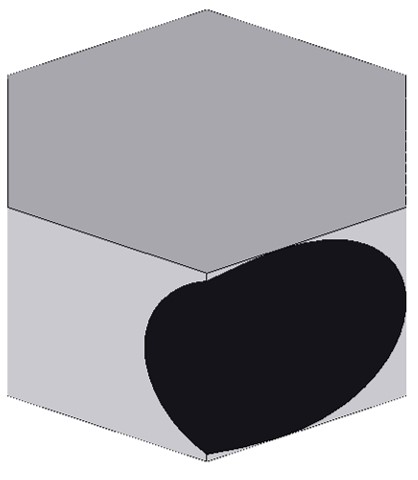}
    }
    \subcaptionbox{}{
        \includegraphics[scale=0.3]{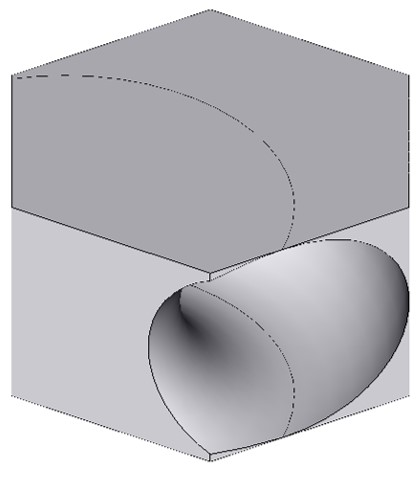}
    }
    \caption{Boolean subtraction result of the prism and torus(prism minus torus)(a) False boolean subtraction result, we can see the faces in the hole of the prism are missing due to the false classification result and the intersection edges on both bottom faces of prism disappear. (b) Right boolean subtraction result.}
    \label{fig:torus and prism boolean result}
\end{figure}

\end{example}

\begin{example}
Considering the cylinder and cone shown in Fig.\ref{fig:exp3-1}, initially, the cylinder's bottom face coincides with the bottom face of the cone, and then the cylinder is rotated by a slight angle around a diameter of the bottom face.

Since the bottom face of the cone now touches the side face of the cylinder only at the two ends of the rotation diameter axis, the intersection entities between the side face of the cylinder and the bottom face of the cone should be two vertices. However, because of the calculation errors, the results are two very short edges. One of them is shown in Fig.\ref{fig:exp3-2}.

\begin{figure}[H]
    \centering
    \includegraphics[scale=0.5]{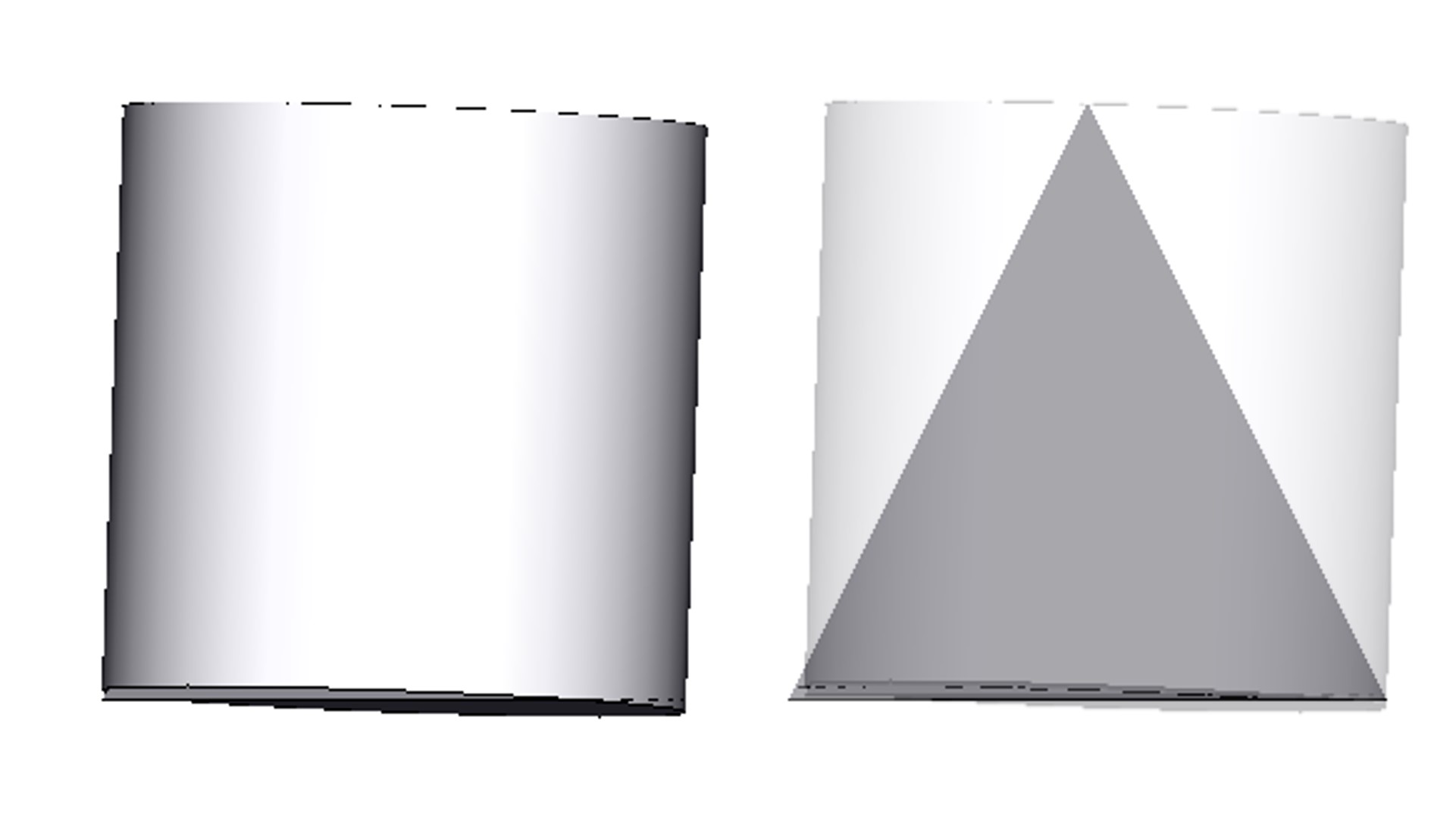}
    \caption{Relative position of the cylinder and cone. Most part of the cone is inside the cylinder.}
    \label{fig:exp3-1}
\end{figure}

These two pretty short edges will not cause noticeable problems in the following boolean steps. However, they will be kept in the final boolean result. In the boolean subtraction result, shown in Fig.\ref{fig:exp3-3}, both short edges are non-manifold edges, an edge associated with more than two faces. The reason is that these short edges are so close to where the faces join that they are falsely regarded as on these faces, which causes significant topological flaws in the resulting body.

\begin{figure}[H]
    \centering
    \includegraphics[scale=0.35]{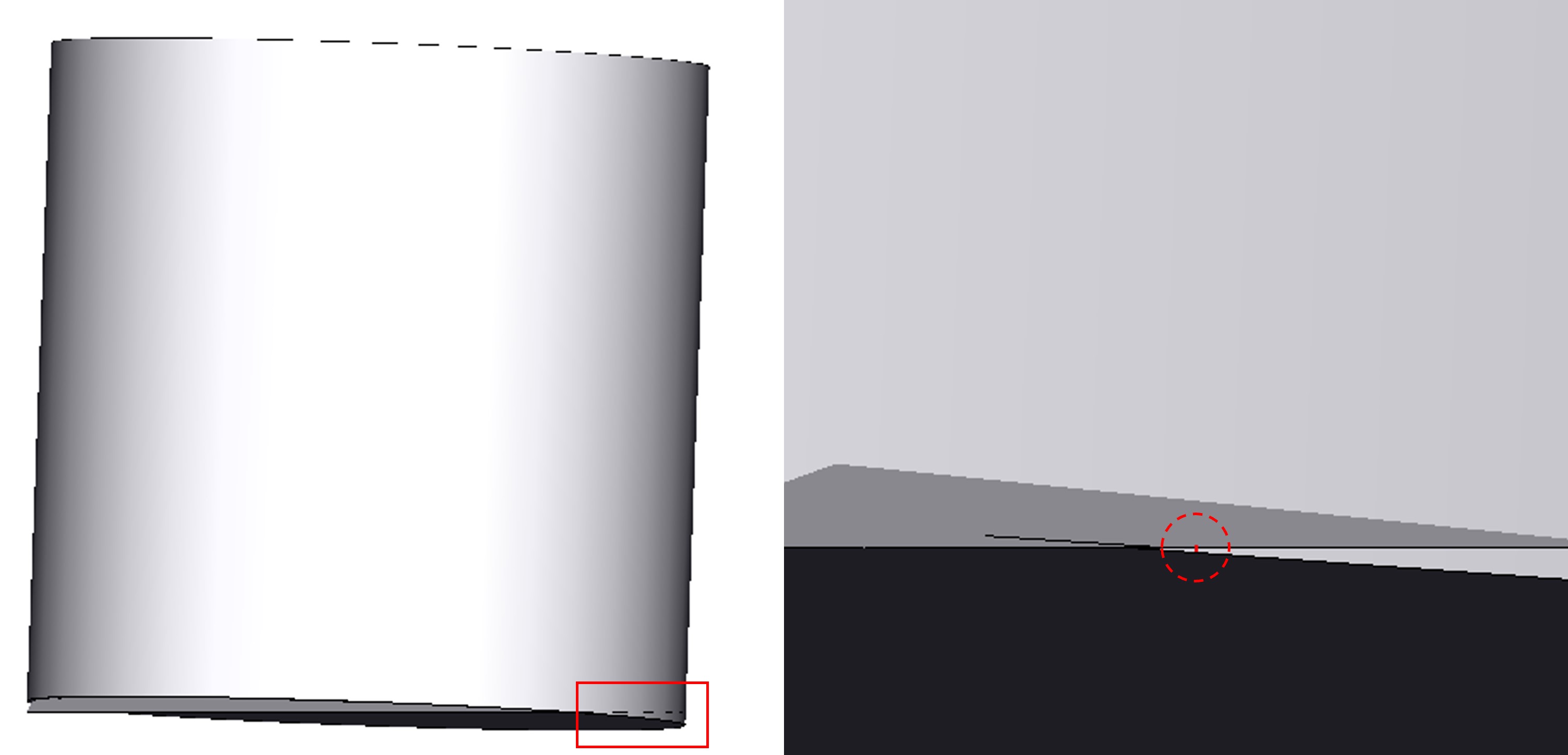}
    \caption{One pretty short intersection edge between the bottom face of the cone and the side face of the cylinder.}
    \label{fig:exp3-2}
\end{figure}

Because these two intersection edges are short enough and isolated in the intersection graph while both ends are on the boundary of faces, they will be detected and removed by the algorithm in \textbf{section \ref{subsubsection:Pretty Short Edges}}. Then the intersection vertices between the cone's bottom face and the cylinder's side face will be added to the intersection graph to replace these two pretty short edges. After repairing, the non-manifold edges in boolean subtraction resulting models are removed, and the whole model becomes 2-manifold again.

\begin{figure}[H]
    \centering
    \includegraphics[scale=0.57]{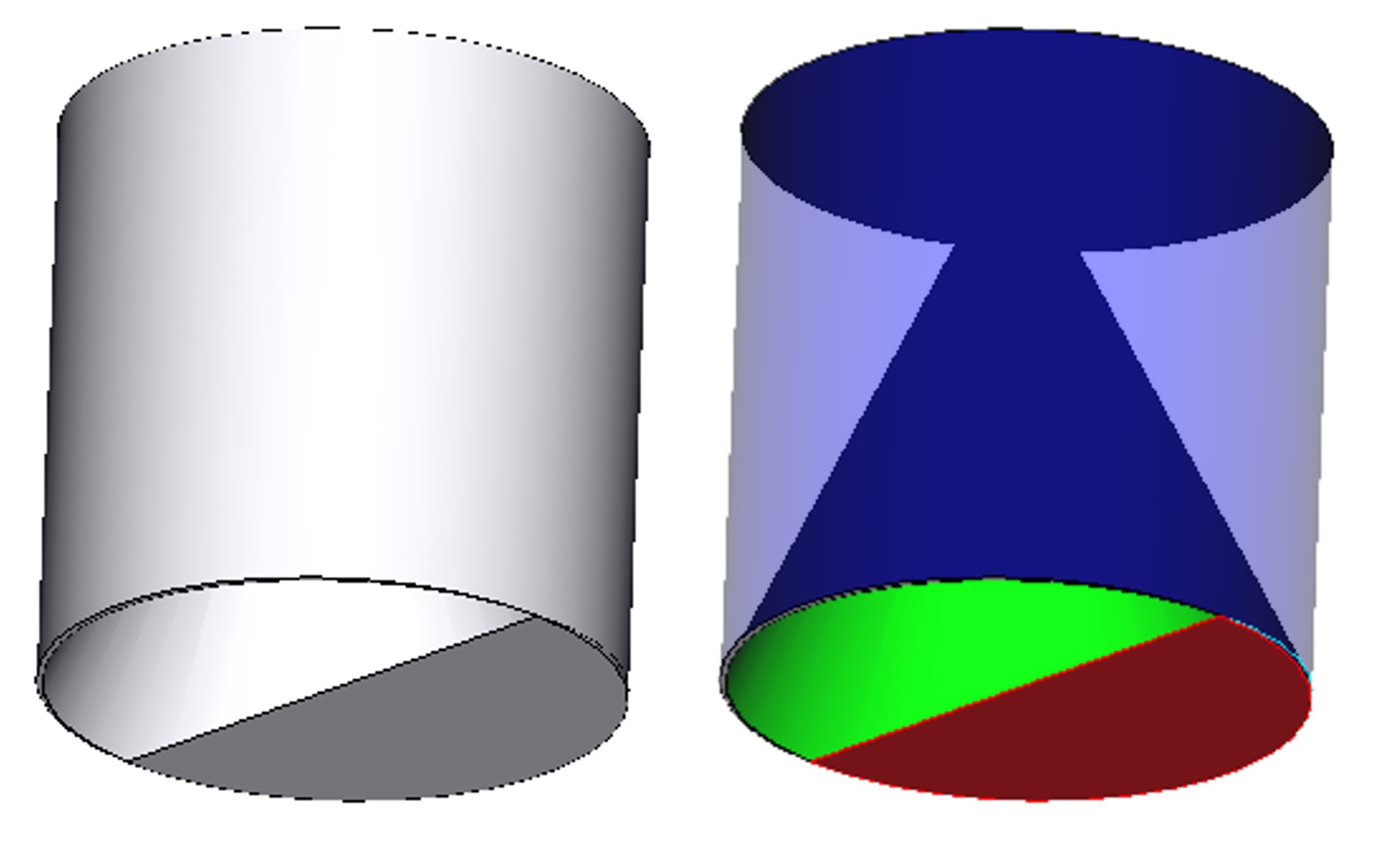}
    \caption{Boolean subtraction result of the cylinder and the cone(cylinder minus cone). Both of the short edges become the boundary edges of all dyed faces. Since the short edges simultaneously lie on four faces (One of the faces is totally blocked by the red one.), the whole model becomes non-manifold.}
    \label{fig:exp3-3}
\end{figure}
\end{example}
\begin{example}
Our method can well preserve the small features in boolean resulting models. Shown in Fig.\ref{fig:Straight ball valve} (a), a straight ball valve is modeled by uniting several simple shapes together. In the dotted box shown in Fig.\ref{fig:Straight ball valve} (b), the position of the green column's bottom face is slightly lower than that of the red cylinder's top face, which makes the green column shallowly embedded in the red cylinder.
    
\begin{figure}[H]
    \centering
    \subcaptionbox{}{
        \includegraphics[scale=0.50]{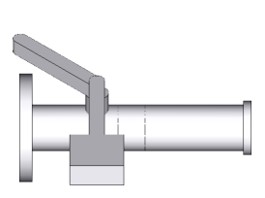}
    }
    \subcaptionbox{}{
        \includegraphics[scale=0.50]{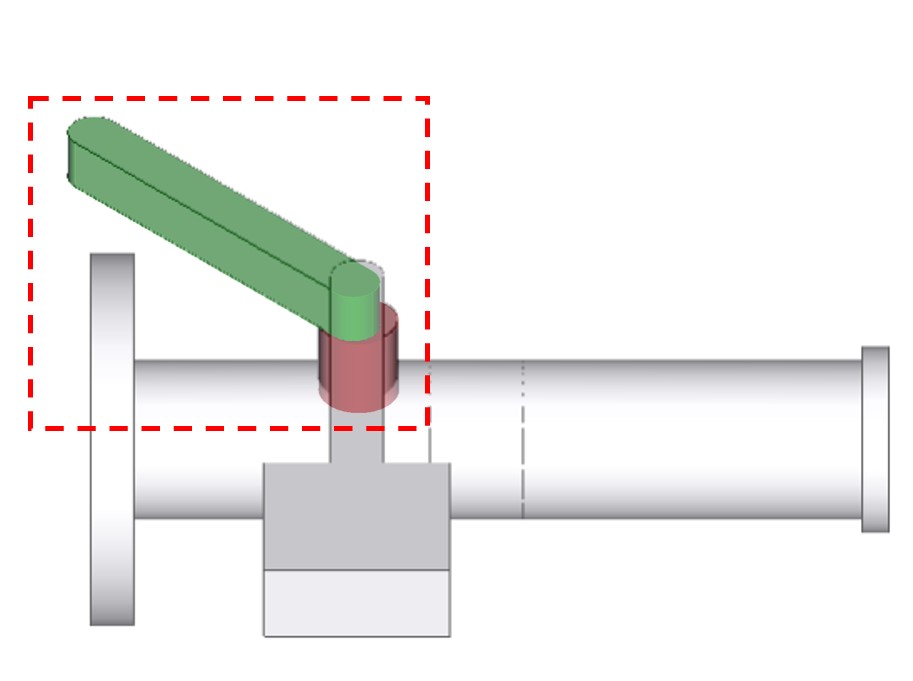}
    }
    \caption{(a) The outlook of the straight ball valve. (b) The colorful part in the dotted box contains a long green column and a red cylinder.}
    \label{fig:Straight ball valve}
\end{figure}

Therefore, in the process of manufacturing this part, there will be a shallow groove on the top of the red cylinder, which is an important feature. However, other repair algorithms which extract features directly from the resulting model or its design history may consider the shallow groove as a result of modeling errors and will remove this small feature by resetting the position of the long column on the top. Without the shallow groove,  the column will only slightly touch the top face of the cylinder and cannot be fixed. Since the constraint is weak, its degrees of freedom are still high, which will cause a lot of trouble in following operations.

\begin{figure}[H]
    \centering
    \includegraphics[scale=0.50]{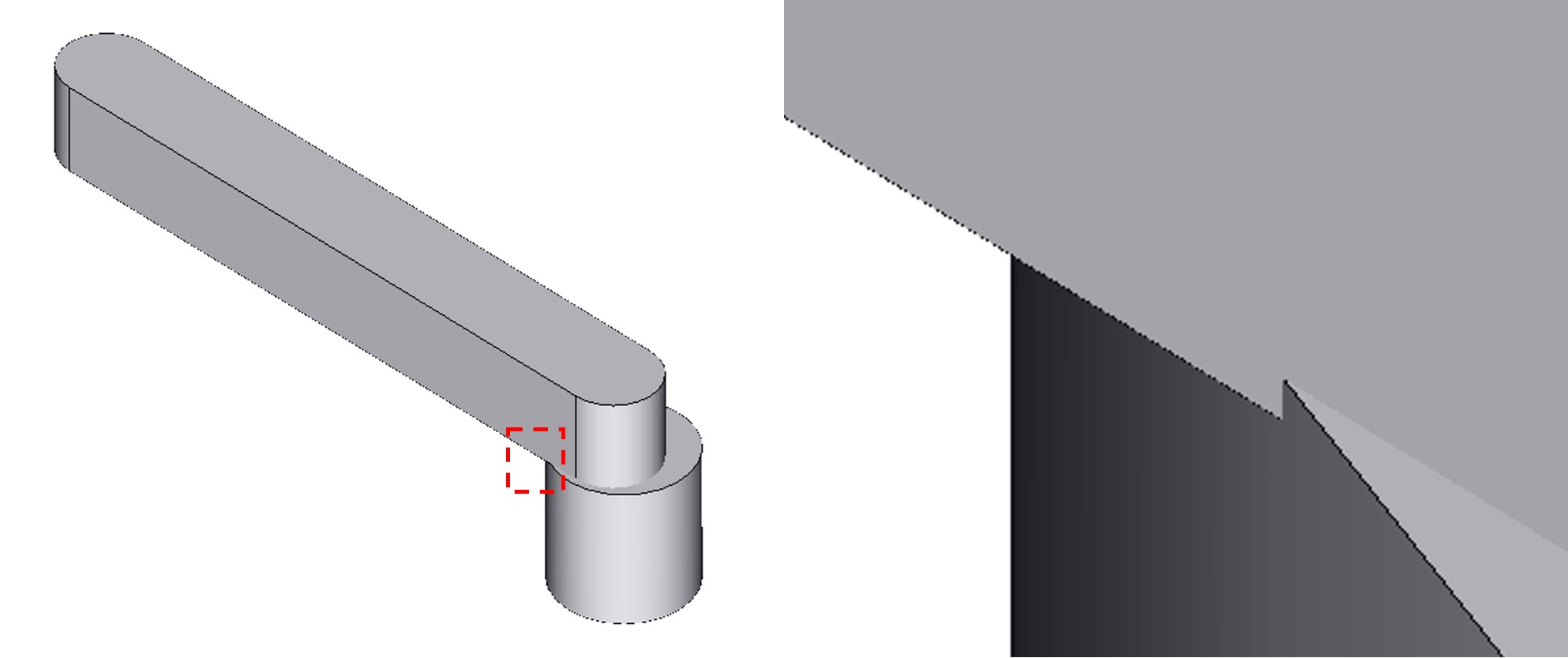}
    \caption{The shallow groove on the top of the cylinder.}
    \label{fig:exp4-3}
\end{figure}

By taking the intersection graph into consideration, our method can well preserve such small features and prevent them from being modified. Shown in Fig.\ref{fig:exp4-4} (b), two very short intersection edges in the intersection graph represent the feature of the shallow groove in the resulting model. Although these two edges are pretty short compared with the size of the whole model, removing them will break the intersection graph, making the area bounded by intersection edges open, which leads to flawed topological structures in resulting models. Therefore, these two edges will be preserved, and so will the shallow groove in the resulting model. With the help of the intersection graph, our method can get more valid topological and geometric information about the model, which directs the inference procedure to make a more reasonable decision and better preserve tiny features in resulting models.

\begin{figure}[H]
    \centering
    \subcaptionbox{}{
    \centering
        \includegraphics[scale=0.50]{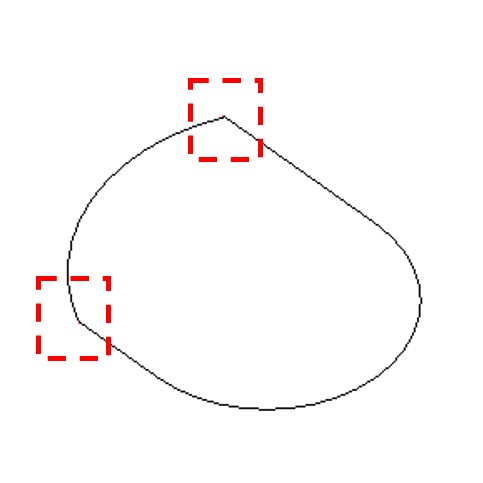}
    }
    \subcaptionbox{}{
    \centering
        \includegraphics[scale=0.50]{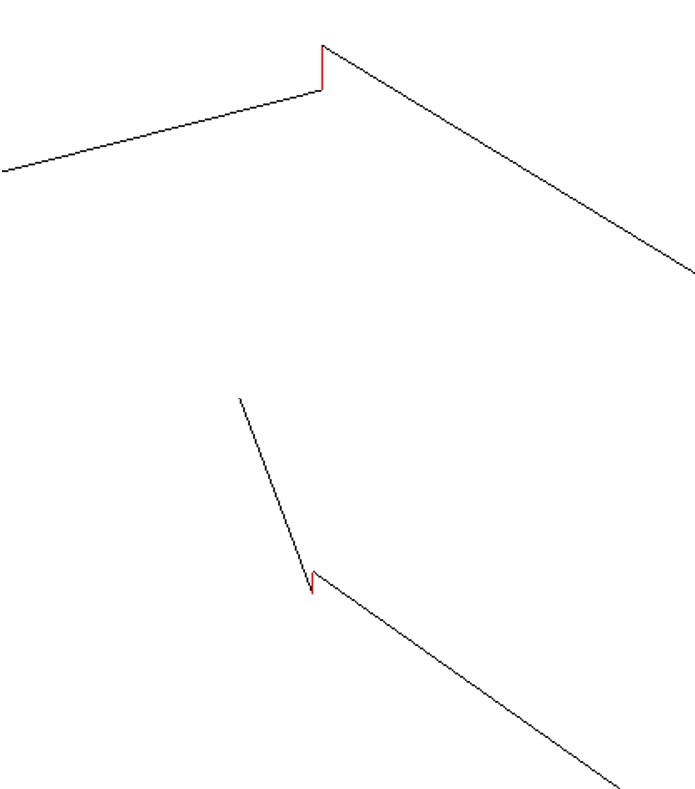}
    }
    \caption{(a) The outlook of the intersection graph of the column and the cylinder. (b) Parts of two dotted boxes after zooming in.}
    \label{fig:exp4-4}
\end{figure}

\end{example}
\section{Conclusion}

This paper is motivated by the illegal topological structures in the resulting models of boolean operations. We found that the illegal topological structures resulted from the inaccurate intersection edges caused by errors. Therefore, we propose a method to detect and repair the flawed intersection edges. Our method uses feature-based tolerance values to direct the inference algorithms to locate flawed intersection edges in a small area. Our inference procedures ensure the repair process will not do unnecessary modifications while correcting the inaccurate intersection edges. And we show the repair effects in different situations in the experiment part. 

For future work, we hope to find a way to calculate a more accurate tolerance value for each intersection edge. And we want to extend our inference algorithms to label a dataset consisting of models with flawed structures,  which will facilitate the use of neural networks to help detect defective models.

\bibliographystyle{elsarticle-num} 
\bibliography{references}  
\end{multicols}

\end{document}